\documentclass[prb,reprint]{revtex4-2}
\usepackage{graphicx,amssymb,amsfonts,amsmath}
\usepackage{color}
\usepackage{mathtools}
\usepackage{latexsym}
\usepackage{graphicx}
\usepackage{float}
\usepackage{dcolumn}
\usepackage{bm}
\usepackage{amssymb}
\usepackage{amsmath}
\usepackage{mathtools}
\usepackage[space]{grffile}
\usepackage{xcolor}
\usepackage{comment}
\usepackage{hyperref}
\hypersetup{
    colorlinks=true,
    linkcolor=blue,
    filecolor=blue,      
    urlcolor=blue,
    citecolor=blue
    }

\newcommand{\insfigsvg}[3]{
\begin{figure}
\centering
\includegraphics[keepaspectratio=true,scale=#2]{#1.jpg}
\caption{#3}
\label{fig:#1}
\end{figure}
}
\begin{document}

\title{Duality defect in a deformed transverse-field Ising model}
\author{Fei Yan$^{\clubsuit}$}
\author{Robert Konik$^{\clubsuit}$}
\author{Aditi Mitra$^{\spadesuit}$}
\affiliation{$^{\clubsuit}$Condensed Matter Physics and Materials Science Division, Brookhaven National Laboratory, Upton, New York 11973, USA\\
$^{\spadesuit}$Center for Quantum
  Phenomena, Department of Physics, New York University, 726 Broadway,
  New York, New York 10003, USA
  }
\begin{abstract}
Physical quantities with long lifetimes have both theoretical significance in the study of quantum many-body systems and practical implications for quantum technologies. In this manuscript, we investigate the roles played by topological defects in the construction of quasi-conserved quantities, using as a prototypical example the Kramers-Wannier duality defect in a deformed 1d quantum transverse field Ising model. We construct the duality defect Hamiltonian in three different ways: half-chain Kramers-Wannier transformation, utilization of techniques in the Ising fusion category, and defect-modified weak integrability breaking deformation. The third method is also applicable for the study of generic integrable defects under weak integrability breaking deformations. We also work out the deformation of defect-modified higher charges in the model and study their slower decay behavior. Furthermore, we consider the corresponding duality defect twisted deformed Floquet transverse field Ising model, and investigate the stability of the isolated zero mode associated with the duality defect in the integrable Floquet Ising model, under such weak integrability breaking deformation.
\end{abstract}
\maketitle

\section{Introduction}

Generic non-integrable closed quantum many-body systems are typically expected to thermalize \cite{Deutsch1991,Srednicki1994,Rigol2008,DAlessio:2015qtq,Gogolin:2015gts,Mori:2017qhg,Mitra_2018}. The total system behaves as a heat bath for its subsystems, leading to the spread of entanglement across the whole system. In the long time limit, expectation values of local observables reach stationary values predicted by statistical ensembles, while the information about the initial state of the system is lost. It is an interesting task to engineer scenarios where quasi-conserved quantities with relatively long lifetimes exist. Not only is this an important topic in quantum many-body theory, but it could also have applications in quantum information storage and processing.

This manuscript concerns the overlap of two setups to construct quasi-conserved quantities with long lifetimes in non-integrable quantum many-body systems. The first setup is a nearly integrable system obtained by turning on weak integrability breaking deformation in an integrable system, where quasi-conserved extensive local charges exist. The second setup is a system with the insertion of defects, which can enable localized long-lived modes. 

In the first setup, the starting point is a quantum system (such as an integrable system) with a set of mutually commuting conserved charges $\{Q_i\}$, consisting of the Hamiltonian as well as higher conserved charges. Turning on a generic deformation, the charges $Q_i$ are not conserved anymore. However, in some cases there could be approximate conserved quantities with long lifetimes. 
Ref.~\cite{Abanin_2017} provided a rigorous theory of such situation in a special class of examples with a separation of energy scales in the Hamiltonian. More recently, the work of \cite{Kurlov2022,Jung2006,Surace:2023wqq} found that for certain special deformations with strength $\lambda$, there exist approximately conserved charges which commute with the Hamiltonian up to corrections of $o(\lambda^2)$. In particular, Ref. \cite{Surace:2023wqq} discussed a systematic analytic approach to construct such quasi-conserved charges, based on the truncation of long-range deformations in integrable models previously studied in \cite{Bargheer:2008jt,Bargheer:2009xy}. The bottom line is that, the system under such special deformations possesses a set of extensive local charges, which mutually commute (and in particular commute with the deformed Hamiltonian) up to corrections of $o(\lambda^2)$. In this sense, such deformations break integrability more weakly than a generic deformation. Moreover, the existence of such quasi-conserved extensive local charges in turn can lead to longer thermalization times. 

The second setup to construct quantities with long lifetimes has a different flavor from the first setup, in the sense that such quantities are associated with boundaries or defects in the system. This direction has been explored a lot for systems with open boundaries. Stable edge modes, termed as {\it strong zero modes}, were constructed in 1d transverse field Ising model, parafermion model, and XYZ model \cite{Kitaev:2000nmw,Fendley:2012vv,Alicea16,FendleyXYZ}. Strong zero modes have lifetimes exponentially long with respect to the system size. Remarkably, such modes are quite robust to perturbations in the sense that {\it almost strong zero modes} with finite but still long lifetimes exist in the corresponding deformed models \cite{Jermyn_2014,Kemp:2017ifs,Else_2017,Carollo2018,Yates20a,Yates20}. Strong edge modes can also exist in periodically-driven Floquet systems \cite{Sen13,Bahri2015,Iadecola2015,KhemaniTC,Potter2017,Potter2018,Wen2017}, which not only host strong zero modes but also strong $\pi$ modes that anti-commute with the Floquet unitary (and more generally strong $2\pi/n$ modes \cite{Moessner2016,Yeh:2023fek}). Floquet strong edge modes are also observed to be robust with respect to perturbations  \cite{Yates19,Yeh:2023uyk,Yeh24a}, in the sense that almost strong Floquet edge modes with long lifetimes exist in deformed Floquet models, even though the bulk heats up to infinite temperature at much shorter time scales.  

Compared with open boundaries, long-lived modes associated with defects are much less studied. Recently a special class of defects, the so-called topological defects, have attracted attention from the fields of condensed matter physics, high energy theory, and quantum information \cite{Verlinde:1988sn,Petkova00,FrohlichFucksRunkelSchweigert,Cobanera:2011wn,Gaiotto:2014kfa,Fendley16,Vidal16,Fendley20,Bhardwaj:2017xup,Buican:2017rxc,Chang:2018iay,Belletete:2018eua,Ji:2019ugf,Komargodski:2020mxz,Gaiotto:2020iye,Thorngren:2021yso,Lootens:2021tet,Roy:2021jus,Freed:2022qnc,Tan22,Cao:2023doz,Seiberg:2023cdc,Sinha:2023hum,Lootens:2023wnl,Seiberg:2024gek,Bhardwaj:2024kvy,ParayilMana:2024txy,Chatterjee:2024ych,Pace:2024tgk,Gorantla:2024ptu,Zhang:2024nbt}. In particular, Ref.~\cite{Fendley16} considered a 1d transverse field Ising (TFI) model on a circular chain with the insertion of a Kramers-Wannier \cite{Kramers1941} duality interface and a domain wall in the coupling space, bi-partitioning the chain into a ferromagnetic phase and a paramagnetic phase. In each $\mathbb{Z}_2$ charge $\pm 1$ sector of the system, there exists a zero mode supported on neighboring sites around the domain wall, which exactly commutes with the Hamiltonian. More recently, Ref.~\cite{Tan22} studied the corresponding problem in a periodically-driven Floquet setting, and found an isolated zero mode localized around the domain wall, exactly commuting with the Floquet unitary. In the special limit where the duality interface is right next to the domain wall thereby shrinking one of the gapped phases to zero size, the zero mode becomes localized around the duality defect. Stability of the zero mode against deformations was probed in \cite{Mitra:2023xdo} for a particular deformed Floquet TFI model. In the presence of the deformation, the zero mode starts to decay. However, the decay mechanism is similar to a reservoir effect, such that for small system sizes the zero mode only partially decays. This can be quantified by the infinite-temperature auto-correlation function of a local operator overlapping with the zero mode, which stabilizes around a finite plateau at infinite time.

In this manuscript, we merge the ideas from two setups to construct physical quantities with long lifetimes. Concretely, we investigate behaviors of defect-modified quantum systems under weak integrability breaking deformations. On the one hand, starting from a defect-modified integrable system, we demonstrate the procedure to construct weakly deformed defect-modified extensive local charges, and study their decay behaviors. On the other hand,  we probe the stability of the zero mode localized around the defect, under weak integrability breaking deformations.

Even though our methods apply to more general quantum systems, here we will streamline the ideas using the example of a 1d deformed TFI model in presence of a Kramers-Wannier (KW) duality defect, both in the Hamiltonian setting and in the periodically-driven Floquet setting. First, we derive the deformed defect Hamiltonian given in Eq.~(\ref{eqn:H_DdTFI1}) for the general case using three different methods: half-chain KW transformation, utilization of techniques from the Ising fusion category, and defect-modified weak integrability breaking deformation. The third method also gives a new framework to study weak integrability breaking in generic integrable systems with integrable defects.  The results agree between these three quite distinct methods. We then work out the deformation of defect-modified higher charges in the model, and study their slower decay behavior under the time evolution by the deformed Hamiltonian. Finally, we move on to the corresponding deformed Floquet TFI model with a duality defect. The zero mode localized around the defect starts to decay in the presence of the deformation. Similar to \cite{Mitra:2023xdo}, for small system sizes the zero mode only decays partially, as corroborated by a finite plateau of the infinite-temperature auto-correlation function for an overlapping local operator. The main novelty of our work is a framework to study weak integrability breaking in integrable systems with integrable defects. Additionally, we generalized the fusion category construction to include next-to-nearest-neighbor projectors suitable for modeling the deformation.

This paper is organized as follows. In Section \ref{sec:defTFI} we give an overview of the KW self-dual deformed TFI model (without any defect insertion) first studied in \cite{Rahmani-Zhu-Franz-Affleck}, and set up some notation conventions. In Section \ref{sec:KWdefectsummary}, we describe the duality defect Hamiltonian in this deformed TFI model and consider the fusion of two duality defects. In Section \ref{sec:halfKW} and Section \ref{sec:fusionKW}, we derive the duality defect Hamiltonian using half-chain KW transformation and techniques from fusion categories respectively. In Section \ref{sec:defectweakintebreaking}, we give a third derivation of the duality defect Hamiltonian using perspectives from weak integrability breaking, and study the slower decay of a deformed higher extensive local charge. In Section \ref{sec:FloquetdTFIdefect}, we turn to the corresponding deformed Floquet TFI model with a duality defect, and probe the stability of the localized zero mode against the deformation. Finally, we conclude in Section \ref{sec:conclusions} with some outlook.

\section{Transverse field Ising model with Kramers-Wannier self-dual deformation} \label{sec:defTFI}

The 1d TFI model can be deformed in different ways that are self-dual under the Kramers-Wannier transformation \cite{Rahmani-Zhu-Franz-Affleck,OBrien:2017wmx}. Here we consider the model first studied in \cite{Rahmani-Zhu-Franz-Affleck}. Our setup is a chain of $L$ sites with site-labels $j=1,...,L$.  
The Hamiltonian for the deformed TFI model with periodic boundary condition is given by
\begin{equation}\label{eqn:HdTFI}
\begin{split}
H_{\text{dTFI}} =&H_{\text{TFI}}+ H_\lambda ~, \text{with}\\
 H_{\text{TFI}}=&-J \sum\limits_{j=1}^{L} Z_{j}Z_{j+1}-g\sum\limits_{j=1}^{L} X_{j}~, \\
H_\lambda=&- \lambda\sum\limits_{j=1}^{L} Z_{j}Z_{j+2}- \lambda\sum\limits_{j=1}^{L}X_{j}X_{j+1} ~.
\end{split}
\end{equation}
Here site $L+1$ is identified with site $1$ and $Z,X$ are Pauli operators acting on the physical spins. The critical point of TFI model is given by $J=g,~ \lambda =0$. Starting from the critical point, turning on $ H_\lambda$ corresponds to an irrelevant deformation from the critical point, which at leading order corresponds to the $T\bar{T}$ deformation \cite{Smirnov:2016lqw,Cavaglia:2016oda} of the continuum Ising conformal field theory. 

The deformation by $H_\lambda$ preserves the $\mathbb{Z}_2$ spin-flip symmetry of the TFI model, generated by 
\begin{equation}\label{eqn:Z2QTFI}
Q_{\mathbb{Z}_2}=\prod\limits_{j=1}^{L}X_{j}~.
\end{equation}
Additionally, $H_\lambda$ is ``invariant" under the Kramers-Wannier (KW) transformation \cite{Kramers1941}, up to a shift by half physical lattice unit, and up to certain global boundary terms. 

For conventional KW transformation, the dual spins are placed on the links, and the dual Pauli operators are given by (using the convention that the string operators are attached to the first site)
\begin{equation}
\begin{split}
\widetilde{X}_{j+\frac{1}{2}}&=Z_{j}Z_{j+1},~ \\
\widetilde{Z}_{j+\frac{1}{2}} &=\prod\limits_{i=1}^{j} X_i, ~~~ j=1,...,L
\end{split}
\end{equation}

After the KW transformation of the TFI model defined on a 1d lattice, we would obtain another TFI model on its dual lattice, i.e. the spins are now placed on the links of the 1d lattice. In the context of this paper though, we would like to remain in the same Hilbert space associated with the original lattice. To achieve this, we perform a half-site translation to the right, and end up with the following Hamiltonian
\begin{equation}
\widetilde{H}_{\text{TFI}}=-J\sum\limits_{j=1}^{L}\widetilde{X}_{j}-g\sum\limits_{j=2}^{L}\widetilde{Z}_{j}\widetilde{Z}_{j+1}-g Q_{\mathbb{Z}_2}\widetilde{Z}_{1}\widetilde{Z}_2~,
\end{equation}
where $Q_{\mathbb{Z}_2}$ is the charge operator for the spin-flip symmetry as defined in Eq.~(\ref{eqn:Z2QTFI}). Therefore up to the boundary term, we obtain a dual TFI model where $J$ and $g$ are exchanged. If the original TFI model is in the ferromagnetic (paramagnetic) phase, the dual TFI model is then in the paramagnetic (ferromagnetic) phase. 

Similarly, the deformation $H_\lambda$ transforms into
\begin{equation}
\begin{split}
\widetilde{H}_\lambda =& -\lambda\sum\limits_{j=1}^L\widetilde{X}_{j}\widetilde{X}_{j+1}-\lambda\sum\limits_{j=2}^{L-1}\widetilde{Z}_j\widetilde{Z}_{j+2}\\
&-\lambda Q_{\mathbb{Z}_2}\widetilde{Z}_{L}\widetilde{Z}_2-\lambda Q_{\mathbb{Z}_2}\widetilde{Z}_1\widetilde{Z}_3 ~.
\end{split}
\end{equation}
Up to boundary terms, $\widetilde{H}_\lambda$ is equivalent to $H_\lambda$. In this sense, the deformation $H_\lambda$ is locally self-dual under the KW transformation.

\section{Duality defect in the self-dual deformed TFI model}\label{sec:KWdefectsummary}

Due to the KW self-dual nature of the deformation $H_\lambda$, we can construct a duality defect in this deformed TFI model. In later Sections, we will describe three different ways to do so. In Section \ref{sec:halfKW}, we construct the duality defect via half-space Kramers-Wannier transformation. In Section \ref{sec:fusionKW}, we illustrate the duality defect construction using techniques from fusion categories, by generalizing previous constructions to next-to-nearest-neighbor projectors that model the deformed theory. In Section \ref{sec:defectweakintebreaking}, we provide another novel way of deriving the duality defect Hamiltonian, as a special kind of weak integrability breaking deformation where the conserved charges in the undeformed model are modified by the defect. The third method gives a new framework to study weak integrability breaking in generic integrable systems with integrable defects, and provides a novel lattice realization of $T\overline{T}$ deformation in the presence of defects. The fact that the results agree between these three different methods serves as a non-trivial check for each individual construction. 

Before we go into details of these different methods, in this Section, we will first describe the resulting duality defect Hamiltonian and demonstrate fusion properties of duality defects. 

The duality defect Hamiltonian for the deformed TFI model on a periodic chain is given by
\begin{equation}\label{eqn:H_DdTFI1}
\begin{split}
H_{D,\text{dTFI}}&=-J\sum\limits_{\substack{j=1,\\ {\color{blue}j\neq i_0}}}^L Z_j Z_{j+1}-g\sum\limits_{\substack{j=1,\\ {\color{blue}j\neq i_0+1}}}^L X_j\\
&-\lambda \sum\limits_{\substack{j=1,\\ {\color{blue}j\neq i_0-1,i_0}}}^L Z_j Z_{j+2}-\lambda \sum\limits_{\substack{j=1,\\ {\color{blue}j\neq i_0,i_0+1}}}^L X_j X_{j+1}\\
&{\color{purple}-J Z_{i_0} X_{i_0+1} -\lambda Z_{i_0-1}X_{i_0+1}}\\
&{\color{purple}-\lambda X_{i_0} Z_{i_0+1} Z_{i_0+2}-\lambda Z_{i_0}X_{i_0+1}X_{i_0+2}}~.
\end{split}
\end{equation}
In particular, the duality defect is introduced by local modifications between sites $i_0-1$ and $i_0+2$, involving both the deletion (in blue) of certain terms in the original model (\ref{eqn:HdTFI}), as well as the addition (in red) of local defect terms. When $J=g$, namely when the undeformed model is the critical TFI model, the Hamiltonian in Eq.~(\ref{eqn:H_DdTFI1}) describes insertion of a KW duality defect in the self-dual deformed critical TFI model, which has also recently appeared in \cite{Seiberg:2024gek}. For the generic cases of $J\neq g$, the Hamiltonian in 
Eq.~\eqref{eqn:H_DdTFI1} can be viewed as describing a KW duality interface together with a domain wall (in the coupling space) inserted by the interface, where across the domain wall the couplings $J$ and $g$ get exchanged while $\lambda$ remains the same. This can also be viewed as a special case of the following configuration: initially we have a KW duality interface and a domain wall inserted at a different location of the circular chain, separating between the deformed paramagnetic phase and ferromagnetic phase. The Hamiltonian in Eq.~(\ref{eqn:H_DdTFI1}) describes the limiting case where one of the two gapped phases is shrunk to zero size.

The model described by Eq. (\ref{eqn:H_DdTFI1}) has a $\mathbb{Z}_2$ symmetry, where the $\mathbb{Z}_2$ charge operator, denoted as $\Omega$, is modified from the $\mathbb{Z}_2$ charge operator in the no-defect TFI model.
\begin{equation}\label{eqn:Z2QTFIdefect}
\begin{split}
\Omega &= \text{i}Z_{i_0+1}Q_{\mathbb{Z}_2}\\
&= - X_1\dots X_{i_0} Y_{i_0+1} X_{i_0+2}\dots X_L~.
\end{split}
\end{equation}

The case of $\lambda=0$, namely the duality defect twisted undeformed TFI model, is exactly solvable by performing a Jordan-Wigner transformation. In particular, in each $\mathbb{Z}_2$ charge sector labeled by $\Omega = \pm 1$, there exists an isolated Majorana zero mode \cite{Fendley16}.  When $J\neq g$, the Majorana zero mode is localized around the domain wall location. When $J=g$, the Majorana zero mode is delocalized along the whole chain. Moreover, this isolated zero mode persists in the corresponding Floquet setup \cite{Tan22}. One of our motivations for this work is to study the fate of the localized zero mode in Floquet TFI model under the KW self-dual deformation, which we will describe in more details in Section \ref{sec:FloquetdTFIdefect}.

\subsection{Fusion of two duality defects}

The KW duality interface is topological, which in particular implies that, there exists a locally-supported unitary which can move the location of the interface \cite{Fendley16,Vidal16,Mitra:2023xdo,Seiberg:2023cdc,Seifnashri:2023dpa,Sinha:2023hum,Seiberg:2024gek}. For ease of discussions, we will consider the case of $J=g$ in the deformed duality defect Hamiltonian in Eq. (\ref{eqn:H_DdTFI1}), which describes the self-dual deformation of the critical TFI model. We  denote this Hamiltonian as $H_{D,i_0}$ to also record the defect location, keeping in mind that $H_{D,i_0}$ involves local modifications between sites $i_0-1$ and $i_0+2$. Then the following unitary
\begin{equation}\label{eqn:shift1}
U_{i_0} : = \text{H}_{i_0}\text{CZ}_{i_0,i_0+1} 
\end{equation}
consisting of a Hadamard gate and a controlled-Z gate, would satisfy
\begin{equation}
U_{i_0} H_{D,i_0} U^\dagger_{i_0} = H_{D,i_0-1}~.
\end{equation}
Because the two defect Hamiltonians $H_{D,i_0}$ and $H_{D,i_0-1}$ are related by a local unitary transformation, physical observables are ignorant of the defect location as long as they are away from the defect sites. Notice that the unitary in Eq. (\ref{eqn:shift1}) is the same kind of unitary which can move the defect location in the undeformed critical TFI model \cite{Fendley16,Vidal16}.

There are three topological defects in the KW self-dual deformed TFI model: the identity defect, the $\mathbb{Z}_2$ defect denoted as $\eta$, and the KW duality defect denoted as $D$. The $\mathbb{Z}_2$ defect $\eta$ is implemented by imposing the anti-periodic boundary condition on the circular chain, concretely its defect Hamiltonian is given by
\begin{equation}
\begin{split}\label{eqn:etadefectham}
H_{\eta,i_0}&= -J \sum\limits_{\substack{j=1,\\j\neq i_0}}^L Z_j Z_{j+1} - J\sum\limits_{j=1}^L X_j \\&-\lambda \sum\limits_{\substack{j=1,\\j\neq i_0,i_0-1}}^L Z_j Z_{j+2} -\lambda\sum\limits_{j=1}^L X_j X_{j+1}\\
&{\color{red}+}J Z_{i_0} Z_{i_0+1}
{\color{red}+}\lambda Z_{i_0-1}Z_{i_0+1} {\color{red}+}\lambda Z_{i_0}Z_{i_0+2}~.
\end{split}
\end{equation}

The three topological defects satisfy a non-trivial fusion relation in the Ising fusion category:
\begin{equation}\label{eqn:isingfusion}
D \times D = I +\eta~.
\end{equation}
In the following, we will see the manifestation of Eq. (\ref{eqn:isingfusion}) on the lattice. The idea is to start with the self-dual deformed TFI model with two duality defect insertions, where the local supports for the two defects do not overlap. For example, consider the situation where one of the defects involves local modifications between sites $i_0-4$ and $i_0-1$ while the other defect involves local modifications between sites $i_0-1$ and $i_0+2$ of the Hamiltonian. We can then move the two defects on top of each other using the unitary in Eq. (\ref{eqn:shift1}). Denoting the two-defect Hamiltonian as $H_{D,i_0-3,i_0}$, then we have
{\small{
\begin{equation}\label{eqn:fusiontransform}
\begin{split}
&U_{i_0-2}U_{i_0-1}U_{i_0}H_{D,i_0-3,i_0} U^\dagger_{i_0}U^\dagger_{i_0-1}U^\dagger_{i_0-2}\\
=&-J\sum\limits_{\substack{j=1,\\j\neq i_0-3,i_0-2}}^L Z_j Z_{j+1}-J\sum\limits_{\substack{j=1,\\j\neq i_0-2}}^L X_j \\
&-\lambda\sum\limits^L_{\substack{j=1,\\ j\neq i_0-4,i_0-3,i_0-2}}Z_jZ_{j+2}
-\lambda \sum\limits_{\substack{j=1,\\j\neq i_0-3,i_0-2}}^L X_j X_{j+1}\\
&-\lambda X_{i_0-3}X_{i_0-1}-J Z_{i_0-3}{\color{red}Z_{i_0-2}}Z_{i_0-1}\\
&-\lambda Z_{i_0-4}{\color{red}Z_{i_0-2}}Z_{i_0-1}-\lambda Z_{i_0-3}{\color{red}Z_{i_0-2}}Z_{i_0}~.
\end{split}
\end{equation}
}}
In particular, the site $i_0-2$ only enters the transformed Hamiltonian through the last three terms in Eq. (\ref{eqn:fusiontransform}). The Pauli operator $Z_{i_0-2}$ commutes with the transformed Hamiltonian, consequently the whole defect Hilbert space admits a decomposition into the $\pm 1$ eigenspace of $Z_{i_0-2}$. In the $+1$ eigenspace, Eq. (\ref{eqn:fusiontransform}) reduces to the no-defect deformed TFI model, albeit on $L-1$ sites labeled by $1,\dots,i_0-3,i_0-1,\dots,L$. In the $-1$ eigenspace, we obtain the $\eta$-defect Hamiltonian (see Eq. (\ref{eqn:etadefectham})) in the deformed TFI model on $L-1$ sites. This defect Hilbert space decomposition, schematically expressed as
\begin{equation}
\mathcal{H}_{D\times D} = \mathcal{H}_I \oplus \mathcal{H}_{\eta}~,
\end{equation}
is a presentation of the fusion rule in Eq. (\ref{eqn:isingfusion}) on the lattice. The reduction in the number of lattice sites in the process of duality defects fusion results from the intricate relation between KW transformation and lattice translation \cite{Seiberg:2024gek,Seiberg:2023cdc,Cheng:2022sgb}.

\section{Duality defect from half-chain Kramers-Wannier transformation}\label{sec:halfKW}

For systems with a certain global symmetry, one can introduce {\it symmetry defects}, which sometimes are also understood as special boundary conditions twisted by the symmetry action. In particular, on an infinite 1d lattice, a symmetry defect can be introduced by performing the corresponding global symmetry action on half of the lattice. Even though the KW duality defect is not associated with ordinary global symmetries, one can still construct it by performing the KW duality transformation on one side of the defect. In fact, away from the critical point, this would produce a duality interface between two gapped phases. In the following, we carry out this procedure for the deformed TFI model. We remark that, even though our original setup is a 1d circular finite chain, to obtain the local modifications in Eq.~(\ref{eqn:H_DdTFI1}) it suffices to work with an effectively infinite chain and perform the half-chain KW transformation. We further remark that there is another closely related construction via the half-chain gauging of the $\mathbb{Z}_2$ symmetry \cite{Seiberg:2024gek}, see also the cases discussed in \cite{Sinha:2023hum,ParayilMana:2024txy,Chatterjee:2024ych,Lu:2024ytl,Cao:2024qjj}.

For the clarity of the derivation, in this Section we will adopt a slightly different convention, similar to the conventions in \cite{Fendley16,Fendley20,Tan22}. We will enlarge the system to consider a chain of $2L$ sites with site-labels $i=0,\dots, 2L-1$. The physical spins are placed on the odd sites, and on the even sites we put auxiliary spins denoted as $\sigma$, which can be thought as associated with the links on the original $L$-site physical spin chain. We remark that, the introduction of auxiliary spins merely serves the purpose of facilitating the demonstration of KW transformation and the derivation of the duality defect Hamiltonian. Furthermore we take $L$ to be large and ignore the boundary details at the ends. In this notation, the duality defect Hamiltonian for the deformed TFI model is given as 
\begin{equation}
\label{eqn:H_DdTFI}
\begin{split}
&H_{D,\text{dTFI}} = -J\sum\limits_{\substack{i=0,\\{\color{blue}i\neq i_0}}}^{L-1}Z_{2i-1}Z_{2i+1}-g\sum\limits_{\substack{i=0,\\{\color{blue}i\neq i_0}}}^{L-1} X_{2i+1}\\
&-\lambda \sum\limits_{\substack{i=0,\\{\color{blue}i\neq i_0-1,i_0}}}^{L-1}Z_{2i-1}Z_{2i+3}-\lambda \sum\limits_{\substack{i=0,\\{\color{blue}i\neq i_0,i_0+1}}}^{L-1}X_{2i-1}X_{2i+1}\\
&{\color{purple}-J Z_{2i_0-1}X_{2i_0+1}-\lambda Z_{2i_0-3}X_{2i_0+1}}\\
&{\color{purple}-\lambda X_{2i_0-1}Z_{2i_0+1}Z_{2i_0+3}-\lambda Z_{2i_0-1}X_{2i_0+1}X_{2i_0+3}}~ .
\end{split}
\end{equation}

We will derive the above Hamiltonian by performing the KW transformation to the right of the site $2i_0$ in the deformed TFI model. Extra care is needed near the interface around site $2i_0$, as the details there determine the required local modifications in the Hamiltonian. In the following, we work in the language of how the KW transformation acts on states in the Hilbert space.

We first consider KW transformation on the whole chain, similar to \cite{Fendley16,Tan22,Cao:2023doz,Sinha:2023hum}. The total Hilbert space of the $2L$-site chain is a direct sum of two subspaces, each of them isomorphic to a physical Hilbert space with $L$ spins. We denote these two subspaces as $\mathcal{H}_{\text{even}}$ and $\mathcal{H}_{\text{odd}}$ respectively, where a typical tensor-product state in $\mathcal{H}_{\text{odd}}$ ($\mathcal{H}_{\text{even}}$) takes the form of $|\sigma h_1 \sigma h_3 ...\sigma h_{2L-1} \rangle$ ($|h_0 \sigma h_2 \sigma ...  h_{2L-2}\sigma  \rangle$). Here the physical spin-up (spin-down) configuration corresponds to $h=0$ ($h=1$) respectively. We denote as $\widehat{D}$ the operator implementing KW transformation on the whole chain. $\widehat{D}$ maps a state in $\mathcal{H}_{\text{odd}}$ to a state in $\mathcal{H}_{\text{even}}$ and vice versa. Starting from a state in $\mathcal{H}_{\text{odd}}$, we have
\begin{equation}
\widehat{D}|\sigma h_1 ... \sigma h_{2L-1}\rangle =\bigotimes_{r=0}^{L-1}\frac{|0_{2r}\sigma\rangle + (-1)^{h_{2r-1}+h_{2r+1}}|1_{2r}\sigma\rangle}{\sqrt{2}} ,
\end{equation}
where details at two ends of the chain depend on the corresponding boundary conditions. One can also work out the action of $\widehat{D}$ on local operators. Concretely, away from the two ends we have
\begin{equation}
\begin{split}
\widehat{D}X_{2r+1}&=\widetilde{Z}_{2r}\widetilde{Z}_{2r+2}\widehat{D} ~,\\
\widehat{D}Z_{2r-1}Z_{2r+1}&=\widetilde{X}_{2r}\widehat{D} ~ .
\end{split}
\end{equation}

Now we apply the KW transformation only to the right of site $2i_0$. As our ultimate goal here is to construct the duality defect Hamiltonian, we would like to stay within the same sub-Hilbert space, e.g. $\mathcal{H}_{\text{odd}}$. This can be achieved by performing an extra one-site (effectively one-half physical spin site) lattice translation on the half chain where the KW transformation is performed. We denote the operator implementing this combined action as $\widehat{D}_{2i_0}$, then
\begin{equation}\label{eqn:d2i0action}
\begin{split}
&\widehat{D}_{2i_0}|\sigma h_1...\sigma h_{2L-1}\rangle\\
&=|\sigma h_1...\sigma h_{2i_0-1}\rangle
\otimes \frac{|\sigma 0_{2i_0+1}\rangle +(-1)^{h_{2i_0+1}}|\sigma 1_{2i_0+1}\rangle}{\sqrt{2}}\\
& \bigotimes_{r=i_0+1}^{L-1}\frac{|\sigma 0_{2r+1}\rangle + (-1)^{h_{2r-1}+h_{2r+1}}|\sigma 1_{2r+1}\rangle}{\sqrt{2}} ~.
\end{split}
\end{equation}

The action of $\widehat{D}_{2i_0}$ on local operators (away from the boundary site $2L-1$) that appear in the Hamiltonian is then given by
\begin{equation}\label{eqn:halfKW}
\begin{split}
{\color{purple}\widehat{D}_{2i_0}Z_{2i_0-1}Z_{2i_0+1}}&={\color{purple} Z_{2i_0-1}X_{2i_0+1} \widehat{D}_{2i_0}}~ ,\\
 \widehat{D}_{2i_0} Z_{2r-1}Z_{2r+1}  &=X_{2r+1} \widehat{D}_{2i_0} ~ , ~ r\geq i_0+1~,\\
 \widehat{D}_{2i_0} X_{2r+1} &= Z_{2r+1} Z_{2r+3}\widehat{D}_{2i_0}  ~ , ~ r\geq i_0 ~.
\end{split}
\end{equation}
In Eq. (\ref{eqn:halfKW}), the red-colored line follows from the Dirichlet boundary condition at the duality interface, under which the operator $\widehat{D}_{2i_0}$ acts as in Eq.~(\ref{eqn:d2i0action}). The relations in Eq.~(\ref{eqn:halfKW}) are crucial to the local modifications in the duality defect Hamiltonian in 
Eq.~(\ref{eqn:H_DdTFI}) for the deformed TFI model. Concretely, after the action of $\widehat{D}_{2i_0}$, away from the duality interface we obtain again a deformed TFI model, while near the duality interface we have
\begin{equation}\label{eqn:DdTFIrelation}
\begin{split}
\widehat{D}_{2i_0} Z_{2i_0-1}Z_{2i_0+1}& = Z_{2i_0-1} X_{2i_0+1} \widehat{D}_{2i_0} ~ ,\\
\widehat{D}_{2i_0} Z_{2i_0-3}Z_{2i_0+1}& = Z_{2i_0-3} X_{2i_0+1} \widehat{D}_{2i_0} ~ ,\\
\widehat{D}_{2i_0} X_{2i_0-1} X_{2i_0+1} &= X_{2i_0-1} Z_{2i_0+1} Z_{2i_0+3} \widehat{D}_{2i_0} ~ ,\\
\widehat{D}_{2i_0} Z_{2i_0-1}Z_{2i_0+3}& = Z_{2i_0-1} X_{2i_0+1}X_{2i_0+3} \widehat{D}_{2i_0} ~ .
\end{split}
\end{equation}
For example, the last relation in Eq. (\ref{eqn:DdTFIrelation}) follows from
\begin{equation}
\begin{split}
\widehat{D}_{2i_0} Z_{2i_0-1}Z_{2i_0+3}& = \widehat{D}_{2i_0} Z_{2i_0-1}Z_{2i_0+1}Z_{2i_0+1}Z_{2i_0+3}\\
&=Z_{2i_0-1} X_{2i_0+1}\widehat{D}_{2i_0}Z_{2i_0+1}Z_{2i_0+3}\\
&=Z_{2i_0-1} X_{2i_0+1}X_{2i_0+3} \widehat{D}_{2i_0} ~.
\end{split}
\end{equation}
The Pauli operators appearing on the RHS of Eq.~(\ref{eqn:DdTFIrelation}) are exactly the new local defect terms appearing in Eq.~(\ref{eqn:H_DdTFI}). Additionally, the deleted terms from the original no-defect Hamiltonian can also be explained. For example, the transverse field operator $X_{2i_0+1}$ is missing in the defect Hamiltonian, which can be explained as follows: after the action of $\widehat{D}_{2i_0}$, the original $X_{2i_0+1}$ becomes $Z_{2i_0+1}Z_{2i_0+3}$, while there are no other operators that get mapped to $X_{2i_0+1}$. Finally, composing with the domain wall in the coupling space, which exchanges $J$ with $g$ to the right of site $2i_0+1$, we obtain the duality defect Hamiltonian in Eq. (\ref{eqn:H_DdTFI}).

\section{Duality defect from fusion categories}\label{sec:fusionKW}

Certain 1d quantum spin chains can be realized as an anyon chain, where techniques from fusion categories can be used to study the physical quantum spin chain \cite{Kitaev06,Trebst3,Fendley16,Fendley20,Buican:2017rxc}. In this section, we utilize such techniques to construct the KW duality defect in the deformed TFI model, after reviewing how this works for the critical TFI model. Throughout this section, we set $J=g=1$. We remark that, the fusion category formalism mainly serves the purpose of identifying the local terms in the untwisted and duality defect twisted Hamiltonians. Although the coupling strengths of such local terms are not canonically encoded, they can be put in at a later stage. 

As usual in the anyonic chain setup, in this Section we will also adopt the same convention as in Section \ref{sec:halfKW}, namely we work with a $2L$-site chain with physical spins placed on the odd sites. 

\subsection{Duality defect in the critical TFI model}

The Hamiltonian for the 1d quantum critical TFI model can be obtained using data of the Ising fusion category. There are three simple objects in the Ising fusion category, which we denote as $I$, $\epsilon$, and $\sigma$, with the nontrivial fusion rules given by
\begin{equation}
\epsilon^2=I,~ \sigma\epsilon=\epsilon\sigma =\sigma, ~ \sigma^2=I+\epsilon
\end{equation}

States in the Hilbert space are represented by {\it fusion trees}, where the vertical branches are associated with a distinguished simple object, taken to be $\sigma$. The horizontal segments of the fusion tree, corresponding to spin sites, are assigned in an alternating pattern of $\sigma$ and $h\in\{I,\epsilon\}$ in a way that is compatible with the Ising fusion rules. Here the physical spin-up and spin-down states are identified with the simple objects $I$ and $\epsilon$ respectively. We remark that, the full space of fusion trees is a direct sum of two subspaces, differing by whether $\sigma$'s are placed on the even or odd sites. Here we will restrict to the subspace where $\sigma$'s are placed on the even sites and physical spins represented by $h\in\{I,\epsilon\}$ are placed on the odd sites. A generic state in this Hilbert space, $|\sigma h_1\sigma h_3...\sigma h_{2L-1}\rangle$, corresponds to the fusion tree illustrated in Figure \ref{fig:fusiontree}. Note that, as the simple objects in the Ising fusion category are self-dual, we can omit the arrows on the branches of the fusion tree.

\insfigsvg{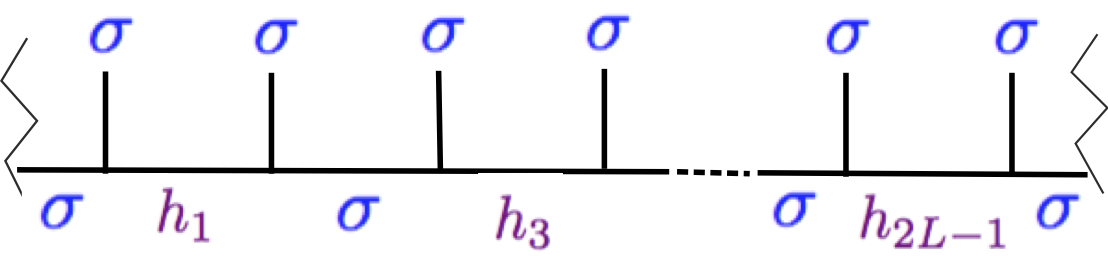}{0.16}{The fusion tree corresponding to the state $|\sigma h_1\sigma h_3...\sigma h_{2L-1}\rangle$, where we have applied the periodic boundary condition. }

In this context, local terms in the Hamiltonian correspond to certain projectors acting on the fusion tree. In particular, the critical TFI Hamiltonian can be written as 
\begin{equation}\label{eqn:HTFIprojector}
H_{\text{TFI}}=-\sum\limits_{i=0}^{L-1}\mathcal{O}^{(1)}_{2i}-\sum\limits_{i=0}^{L-1}\mathcal{O}^{(2)}_{2i+1} ~,
\end{equation}
where $\mathcal{O}^{(1)}_{2i}$ and $\mathcal{O}^{(2)}_{2i+1}$ are local projectors acting on the fusion tree, by spanning an extra horizontal branch labeled by $\epsilon$ between two vertical legs. Diagrammatically this is illustrated in Figure \ref{fig:projector}.

\insfigsvg{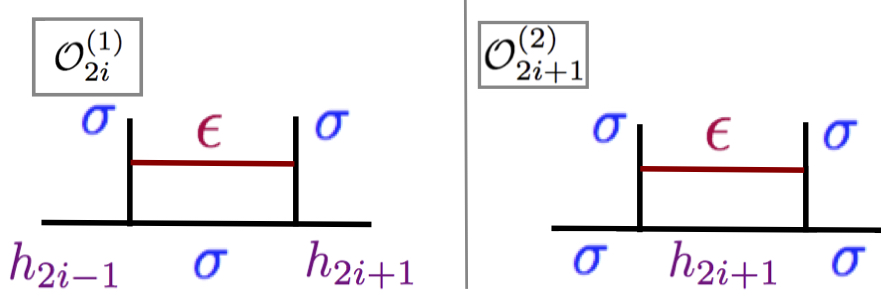}{0.19}{Diagrammatic presentation of the local terms $\mathcal{O}_{2i}^{(1)}$ and $\mathcal{O}_{2i+1}^{(2)}$ in Eq. \eqref{eqn:HTFIprojector} in the context of the fusion tree.}

To explicitly write such local terms in terms of Pauli operators, we evaluate the matrix elements associated with $\mathcal{O}^{(1)}_{2i}$ and $\mathcal{O}^{(2)}_{2i+1}$. For example, the local matrix element $\langle \sigma h'_{2i+1} \sigma | \mathcal{O}^{(2)}_{2i+1} | \sigma h_{2i+1}\sigma \rangle$ corresponds to the diagram illustrated in Figure \ref{fig:pauliX}. Throughout this Section, we will use the abbreviated notation for states where we only keep the sites relevant for the matrix element evaluation. Notice that the bra state corresponds to a fusion tree placed upside down, and the diagram in Figure \ref{fig:pauliX} is produced by contracting the corresponding branches.  We remark that, here we use a normalization such that physical matrix elements are given by the evaluation of diagrams (such as the one in Figure \ref{fig:pauliX}) divided by a constant, where this constant is given by the quantum dimension of $\sigma$ raised to the power of the number of $\sigma$ loops, where  the quantum dimension of $\sigma$ is $\sqrt{2}$.

\insfigsvg{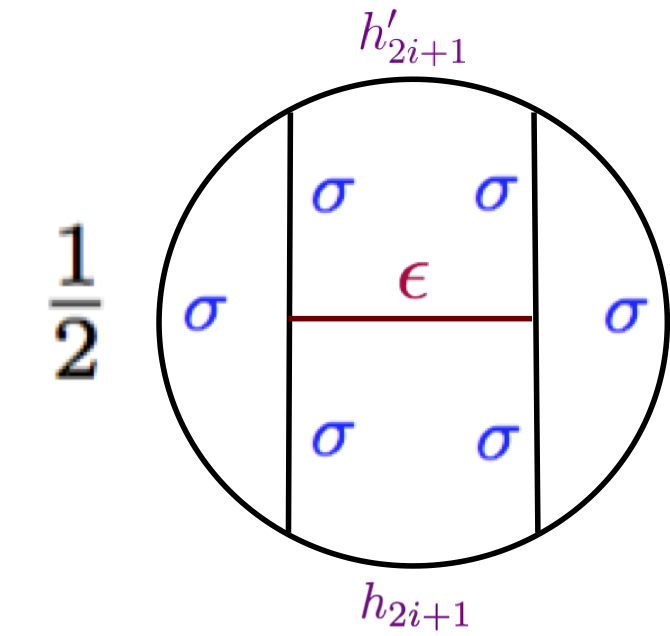}{0.128}{Diagrammatic presentation of the matrix element $\langle \sigma h'_{2i+1} \sigma | \mathcal{O}^{(2)}_{2i+1} | \sigma h_{2i+1}\sigma \rangle$. The factor of $\frac{1}{2}$ comes from the two closed $\sigma$-loops in the diagram.}

\insfigsvg{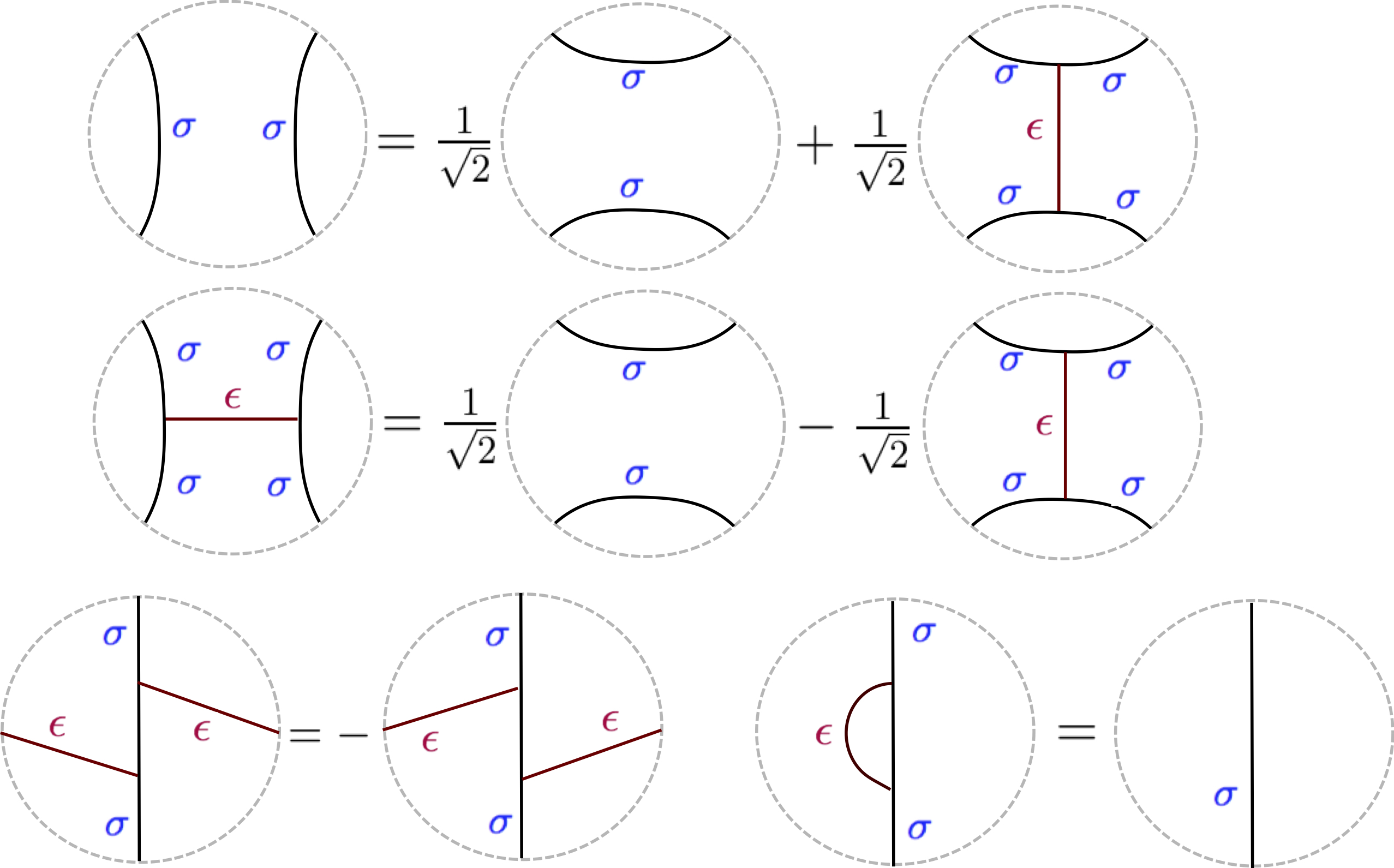}{0.09}{Non-trivial fundamental F-moves in the Ising fusion category.}

\insfigsvg{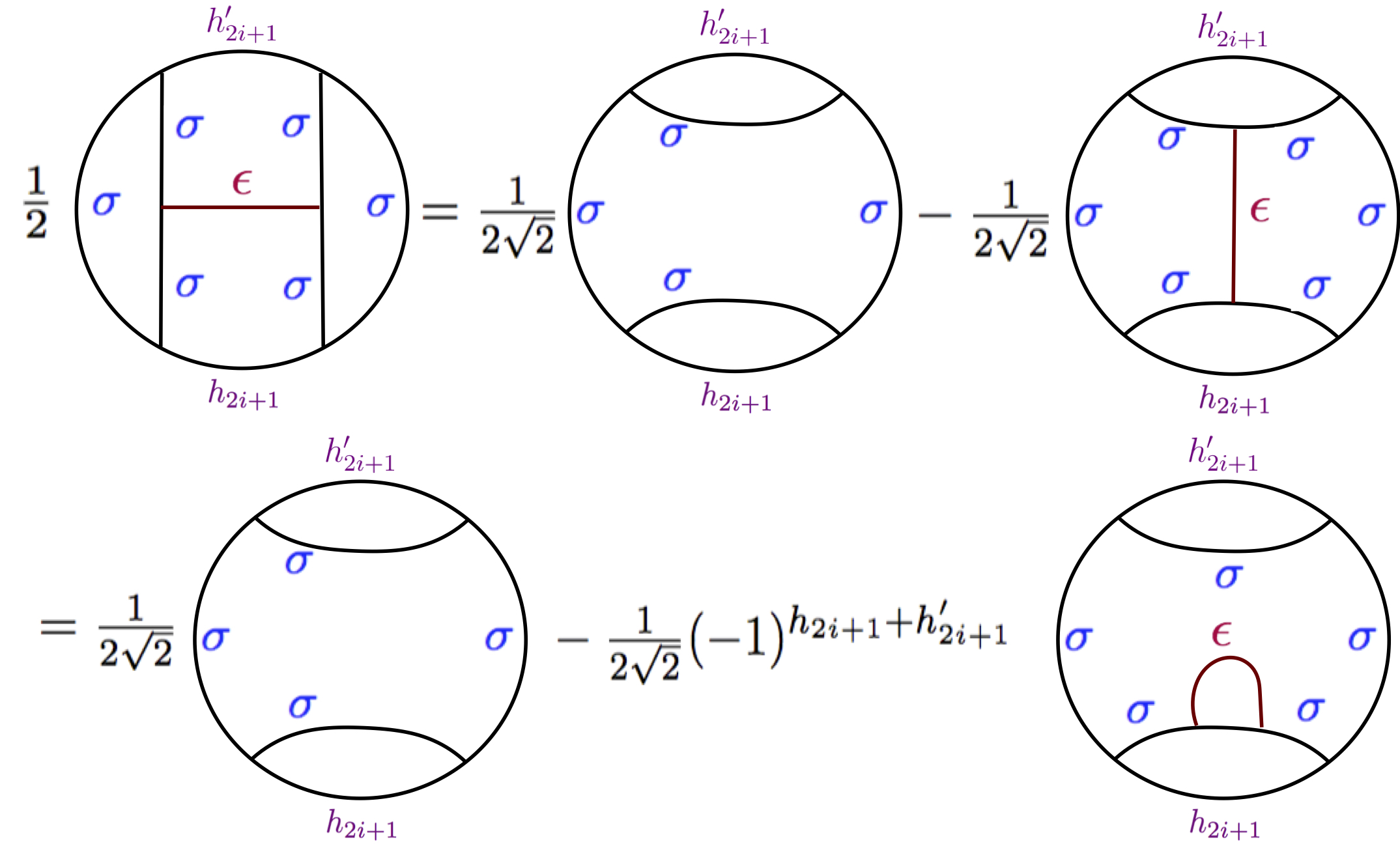}{0.11}{Simplification of the diagram corresponding to the matrix element $\langle \sigma h'_{2i+1} \sigma | \mathcal{O}^{(2)}_{2i+1} | \sigma h_{2i+1}\sigma \rangle$ using F-moves.}

Using the F-symbols of the Ising fusion category \cite{Fendley20,Chang:2018iay}, where we collect the fundamental non-trivial F-moves in Figure \ref{fig:Fmoves}, the diagram in Figure \ref{fig:pauliX} can be simplified and evaluated. This process is illustrated in Figure \ref{fig:pauliXFsymbol}. According to the last row of Figure \ref{fig:pauliXFsymbol} together with the quantum dimension $d_\sigma =\sqrt{2}$, we have
\begin{equation*}
\langle \sigma h'_{2i+1} \sigma | \mathcal{O}^{(2)}_{2i+1} | \sigma h_{2i+1}\sigma \rangle= \frac{1}{2}\left(1-(-1)^{h_{2i+1}+h'_{2i+1}} \right)~.
\end{equation*}
If $h_{2i+1}=h'_{2i+1}$, then the matrix element vanishes. Otherwise, $h_{2i+1}+h'_{2i+1}=1$ and the matrix element evaluates to $1$. Therefore we see that the local term $\mathcal{O}^{(2)}_{2i+1}$ acts as the Pauli operator $X_{2i+1}$. 

For the other type of local term $\mathcal{O}^{(1)}_{2i}$, we can evaluate its matrix element in a similar fashion. This is illustrated in Figure \ref{fig:pauliZZFsymbol}. We have
\begin{equation*}
\langle h_{2i-1}\sigma h_{2i+1}| \mathcal{O}^{(1)}_{2i} | h_{2i-1}\sigma h_{2i+1}\rangle 
=(-1)^{h_{2i-1}+h_{2i+1}} ~ .
\end{equation*}
Therefore $\mathcal{O}^{(1)}_{2i}$ is equivalent to the local two-site operator $Z_{2i-1}Z_{2i+1}$. 

\insfigsvg{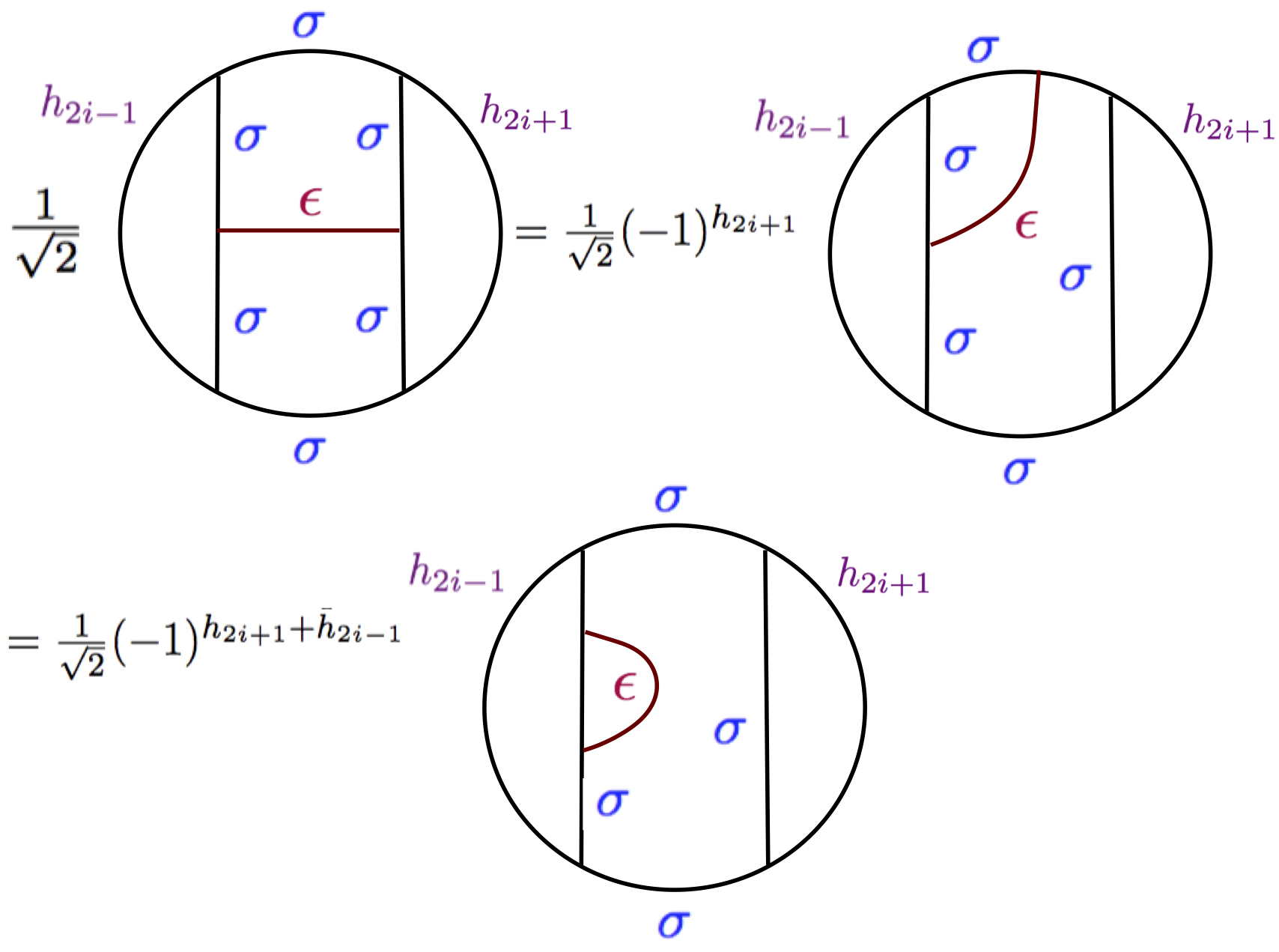}{0.12}{Evaluation of the diagram corresponding to the matrix element $\langle h_{2i-1}\sigma h_{2i+1}| \mathcal{O}^{(1)}_{2i} | h_{2i-1}\sigma h_{2i+1}\rangle$ using F-moves.}

Now that we have recovered the critical TFI Hamiltonian from the fusion category language, we proceed to insert a duality defect, between physical spins at sites $2i_0-1$ and $2i_0+1$. States in the duality defect Hilbert space are represented by fusion trees with one vertical branch inserted from below, as illustrated on the LHS in Figure \ref{fig:fusiontreedualitydefect}.

\insfigsvg{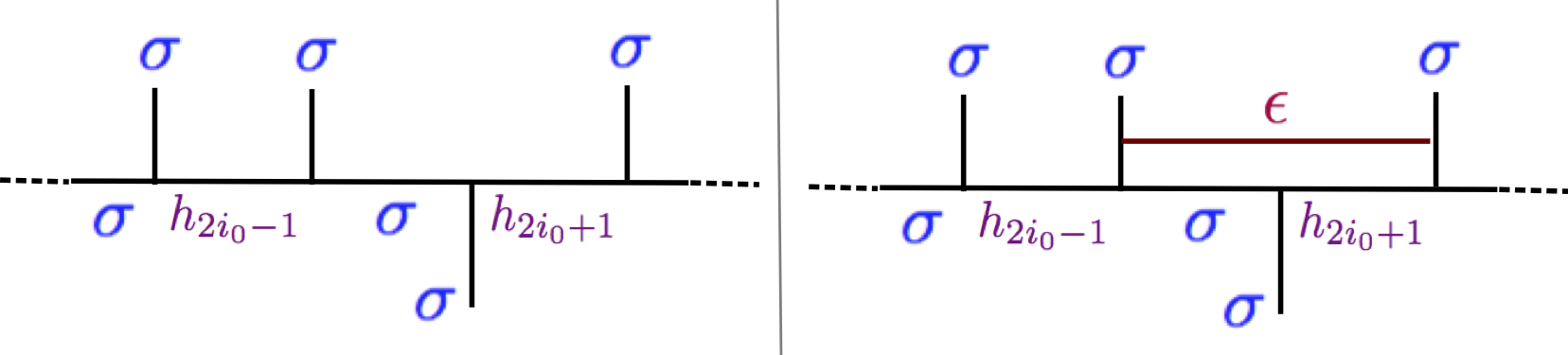}{0.14}{{\it Left}: The fusion tree representing states in the duality defect Hilbert space. {\it Right}: The action of local defect term $\mathcal{O}^{D,\text{TFI}}_{2i_0}$ on the fusion tree.}

\insfigsvg{PauliZX}{0.12}{Diagrammatic presentation for the matrix element $\langle h_{2i_0-1}\sigma h'_{2i_0+1}|\mathcal{O}^{\text{D,TFI}}_{2i_0}|h_{2i_0-1}\sigma h_{2i_0+1} \rangle$.}

The duality defect Hamiltonian in the critical TFI model can be obtained, again by applying projectors on the fusion trees. In particular, there is a new defect local term $\mathcal{O}^{D,\text{TFI}}_{2i_0}$, as illustrated on the RHS of Figure \ref{fig:fusiontreedualitydefect}. Its matrix element is diagrammatically represented in Figure \ref{fig:PauliZX}, which also demonstrated the simplification of such a diagram. In the end, this matrix element evaluates as
\begin{equation}
\begin{split}
&\langle h_{2i_0-1}\sigma h'_{2i_0+1}|\mathcal{O}^{\text{D,TFI}}_{2i_0}|h_{2i_0-1}\sigma h_{2i_0+1}\rangle\\
&= \sum\limits_{x}\frac{1}{2}(-1)^{(h_{2i_0+1}+h'_{2i_0+1})x}(-1)^{x+h_{2i_0-1}}
\end{split}
\end{equation}
Notice that, if $h_{2i_0+1}=h'_{2i_0+1}$, then this matrix element vanishes. Therefore we must have $h_{2i_0+1}+h'_{2i_0+1}=1$, and the matrix element reduces to $(-1)^{h_{2i_0-1}}$, from which we conclude that the local defect term $\mathcal{O}^{D,\text{TFI}}_{2i_0}$ is equivalent to the two-site operator $Z_{2i_0-1}X_{2i_0+1}$. Moreover, from the projector presentation in Figure \ref{fig:fusiontreedualitydefect} we also see that there are no $Z_{2i_0-1}Z_{2i_0+1}$ and $X_{2i_0+1}$ in the Hamiltonian. Thus, we recover the well-known KW duality defect Hamiltonian in the critical TFI model. 

\subsection{Duality defect in the deformed TFI model}

We now consider the self-dual deformation of TFI model given in Eq.~(\ref{eqn:HdTFI}). The deformation $H_\lambda$ can be written in terms of next-to-nearest-neighbor projectors, 
\begin{equation}\label{HDTFIprojector}
H_\lambda =-\lambda \left(\sum\limits_{i=0}^{L-1}P^{(1)}_{2i+1} + \sum\limits_{i=0}^{L-1}P^{(2)}_{2i}\right) ~.
\end{equation}
Here $P^{(1)}_{2i+1}$ and $P^{(2)}_{2i}$ are local operators acting on the fusion tree, whose diagrammatic representations are given in Figure \ref{fig:projector2}.

\insfigsvg{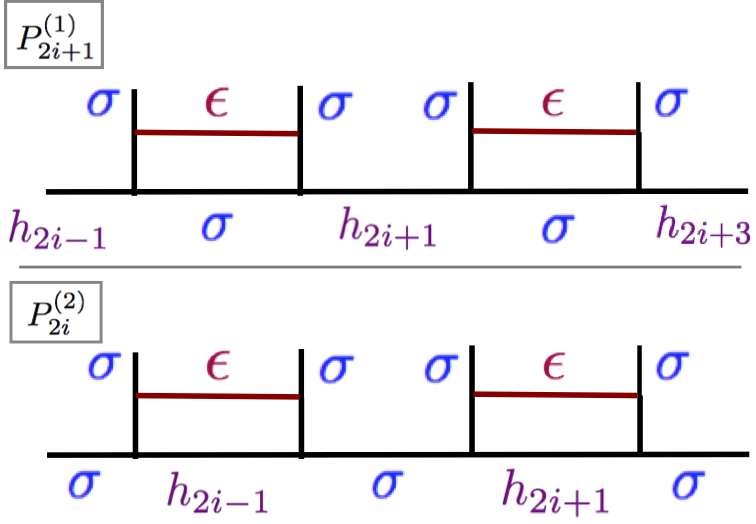}{0.18}{Diagrammatic presentation of the local terms $P^{(1)}_{2i+1}$ and $P^{(2)}_{2i}$ in Eq. \eqref{HDTFIprojector} in the fusion tree context.}

We can express $P^{(1)}_{2i+1}$ and $P^{(2)}_{2i}$ in terms of Pauli operators, by evaluating the diagrammatic representation of their matrix elements. First, we consider the matrix element $\langle h_{2i-1}\sigma h'_{2i+1}\sigma h_{2i+3}|P^{(1)}_{2i+1}|h_{2i-1}\sigma h_{2i+1}\sigma h_{2i+3}\rangle$. If $h'_{2i+1}=h_{2i+1}$, then inspecting the diagrammatic presentation of the matrix element yields that it factorizes as
\begin{equation}
\begin{split}
&\langle h_{2i-1}\sigma h_{2i+1}\sigma h_{2i+3}|P^{(1)}_{2i+1}|h_{2i-1}\sigma h_{2i+1}\sigma h_{2i+3}\rangle\\
&=\langle h_{2i-1}\sigma h_{2i+1}|\mathcal{O}^{(1)}_{2i}|h_{2i-1}\sigma h_{2i+1}\rangle \\
&\times \langle h_{2i+1}\sigma h_{2i+3}|\mathcal{O}^{(1)}_{2i+2}|h_{2i+1}\sigma h_{2i+3}\rangle ~.
\end{split}
\end{equation}
Therefore $P^{(1)}_{2i+1}$ is equivalent to $Z_{2i-1}Z_{2i+1}Z_{2i+1}Z_{2i+3}=Z_{2i-1}Z_{2i+3}$. On the other hand, if $h'_{2i+1}\neq h_{2i+1}$, then the matrix element of $P^{(1)}_{2i+1}$ vanishes. This can be seen from the diagrammatic simplification using F-moves, where in the end the diagram contains a tadpole enforcing the vanishing of the diagram \cite{Chang:2018iay,Fendley16}, as illustrated in Figure \ref{fig:pauliZZZZ}.

\insfigsvg{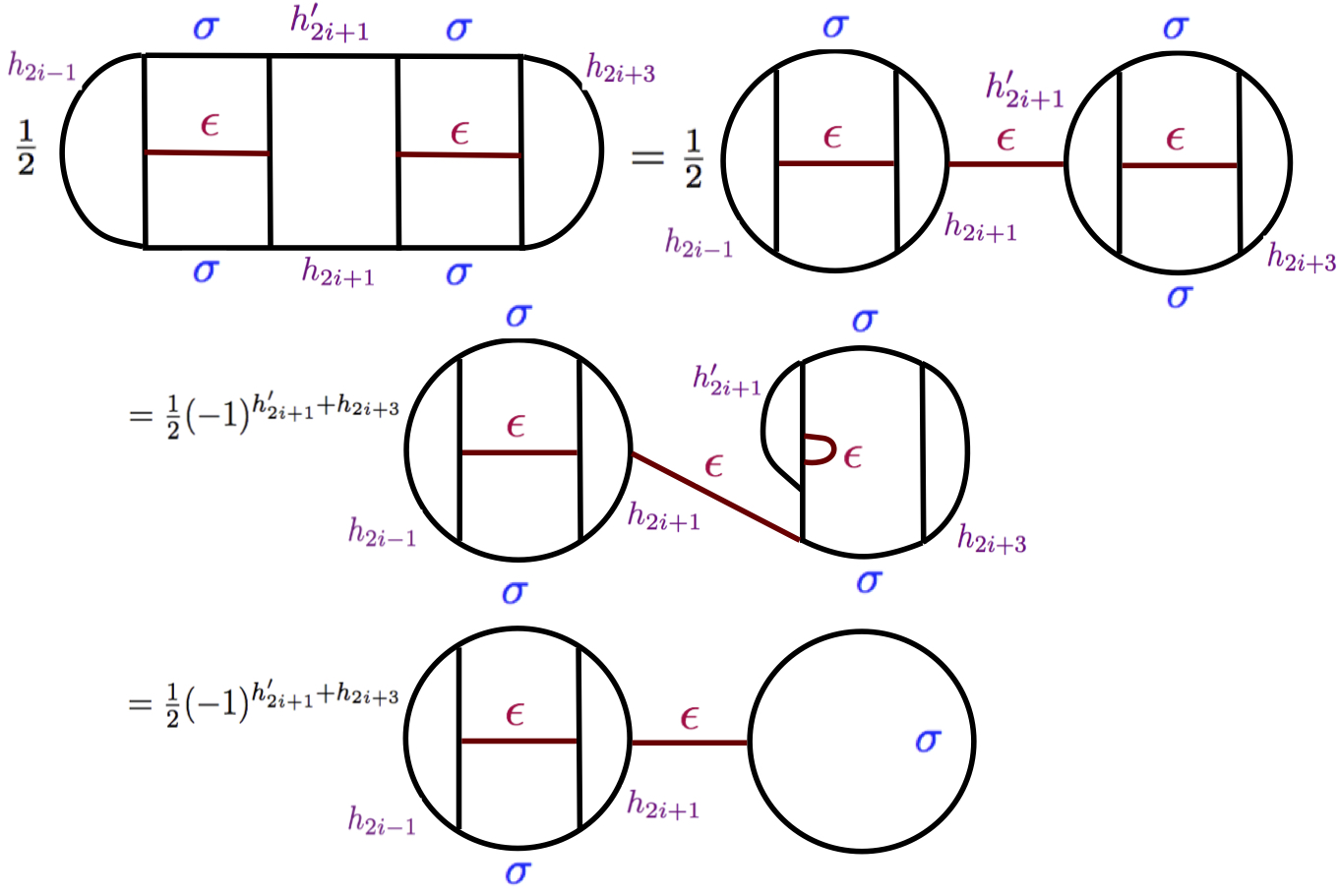}{0.18}{Diagrammatic simplification for the matrix element of $P^{(1)}_{2i+1}$ in Eq. \eqref{HDTFIprojector} when $h'_{2i+1}\neq h_{2i+1}$.}

\insfigsvg{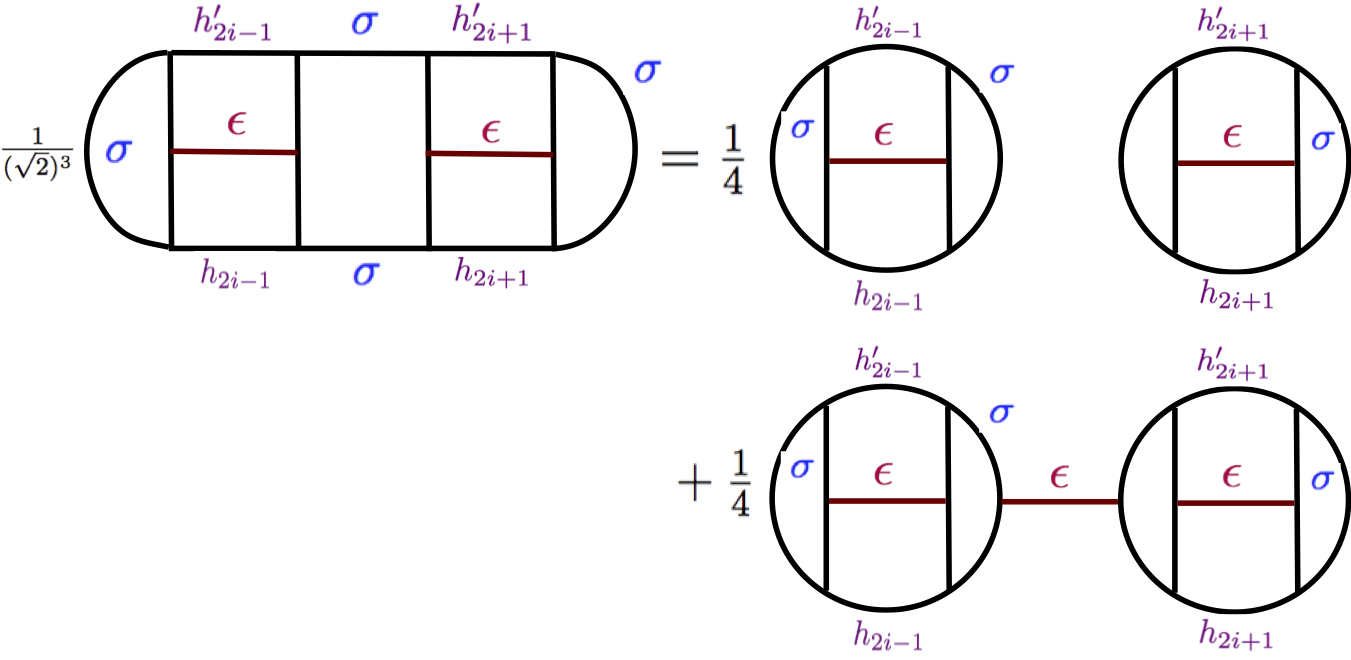}{0.18}{Diagrammatic presentation for the matrix element $\langle \sigma h'_{2i-1}\sigma h'_{2i+1}\sigma |P^{(2)}_{2i}| \sigma h_{2i-1}\sigma h_{2i+1} \sigma \rangle$.}

We now turn to the matrix element of $P^{(2)}_{2i}$, namely $\langle \sigma h'_{2i-1}\sigma h'_{2i+1}\sigma |P^{(2)}_{2i}| \sigma h_{2i-1}\sigma h_{2i+1} \sigma \rangle$, which can be diagrammatically represented as in Figure \ref{fig:pauliXX}. The first term on the RHS of Figure \ref{fig:pauliXX} again corresponds to the following factorization:
\begin{equation*}
\frac{1}{4}\times 2 \langle \sigma h'_{2i-1}\sigma|\mathcal{O}^{(2)}_{2i-1}|\sigma h_{2i-1}\sigma\rangle \times  2 \langle \sigma h'_{2i+1}\sigma|\mathcal{O}^{(2)}_{2i+1}|\sigma h_{2i+1}\sigma\rangle.
\end{equation*}
The second term on the RHS of Figure \ref{fig:pauliXX} again vanishes due to a tadpole, as illustrated in Figure \ref{fig:pauliXX2}. Therefore $P^{(2)}_{2i}$ is equivalent to the two-site Pauli operator $X_{2i-1}X_{2i+1}$.

\insfigsvg{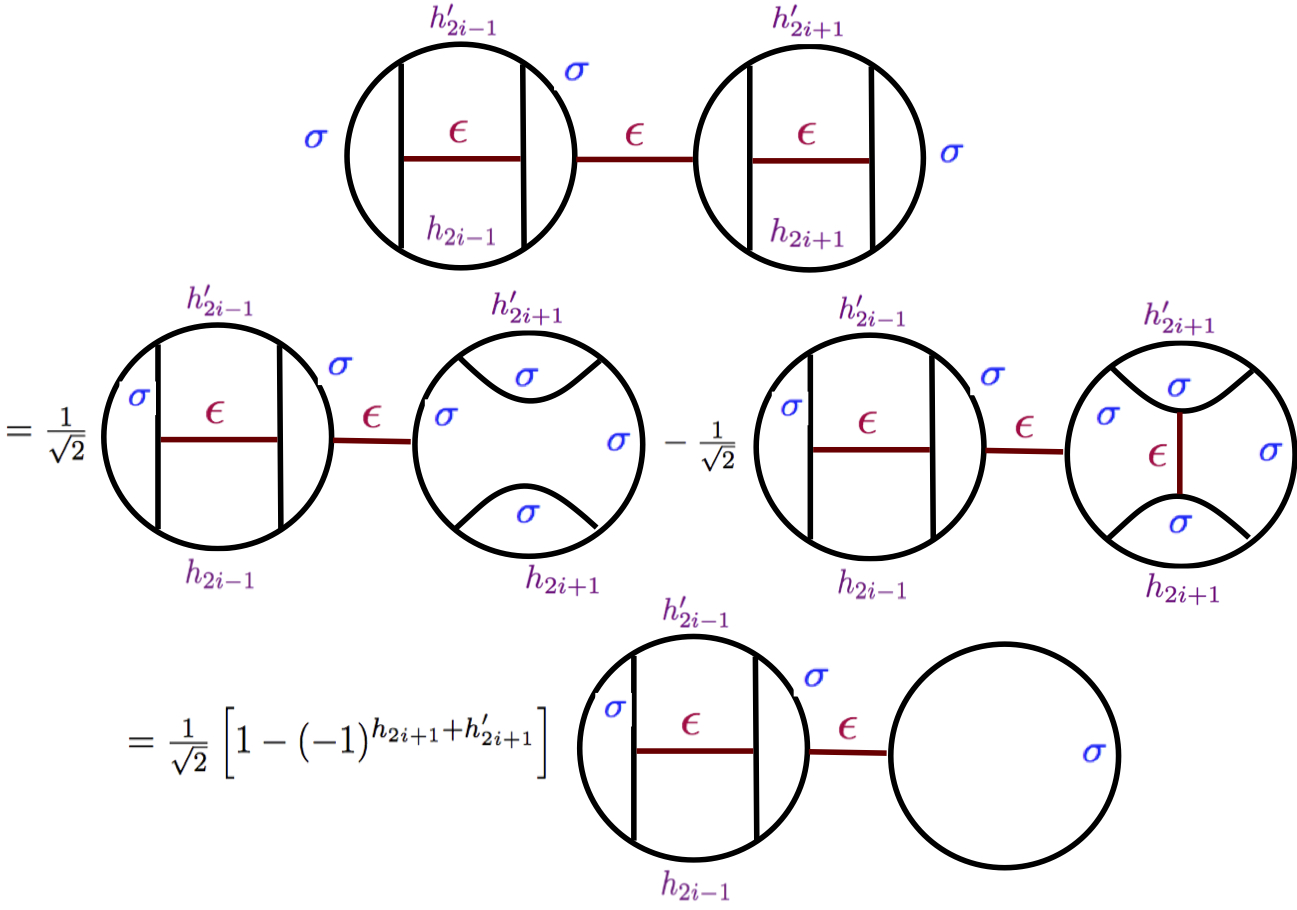}{0.18}{Diagrammatic simplification for the second term on the RHS of Figure \ref{fig:pauliXX}.}

We are now ready to work out the defect Hamiltonian in the deformed TFI model. Other than the local defect term $\mathcal{O}_{2i_0}^{D,\text{TFI}}$ graphically demonstrated on the RHS of Figure \ref{fig:fusiontreedualitydefect}, there are three additional local defect terms. We denote them as $\mathcal{A}_{2i_0}^{\text{dTFI}}$, $\mathcal{B}_{2i_0}^{\text{dTFI}}$, and $\mathcal{C}_{2i_0}^{\text{dTFI}}$. Their actions on the fusion tree are illustrated in Figure \ref{fig:fusiontreedualitydefectdTFI}.

\insfigsvg{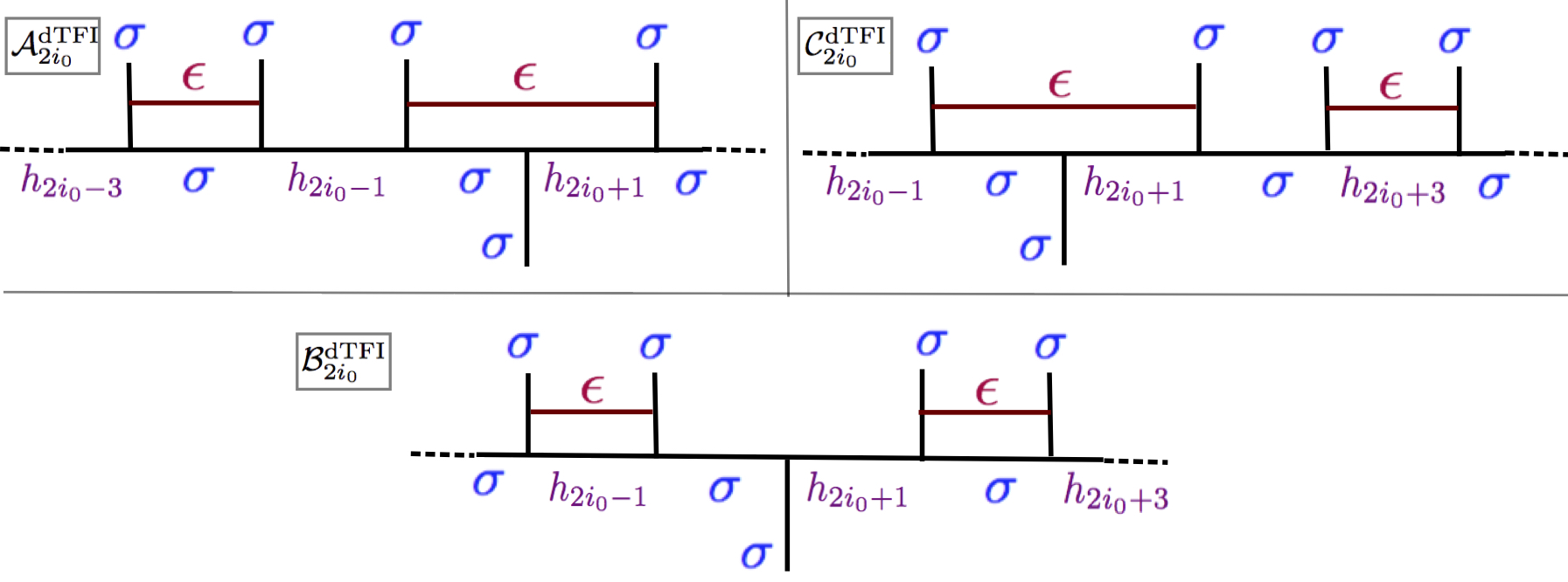}{0.14}{Diagrammatic presentation for the action of defect terms $\mathcal{A}_{2i_0}^{\text{dTFI}}$, $\mathcal{B}_{2i_0}^{\text{dTFI}}$, and $\mathcal{C}_{2i_0}^{\text{dTFI}}$ on the fusion tree.}

We first consider the matrix element $\langle h_{2i_0-3}\sigma h'_{2i_0-1}\sigma h'_{2i_0+1}\sigma| \mathcal{A}_{2i_0}^{\text{dTFI}}|h_{2i_0-3}\sigma h_{2i_0-1}\sigma h_{2i_0+1}\sigma\rangle$. Similar to the analysis of the matrix element for $P^{(1)}_{2i+1}$ (see Figure \ref{fig:pauliZZZZ}), if $h_{2i_0-1}\neq h'_{2i_0-1}$, the matrix element vanishes. Therefore $h_{2i_0-1}=h'_{2i_0-1}$ and the matrix element factorizes. In terms of Pauli operators, $\mathcal{A}_{2i_0}^{\text{dTFI}}$ is equivalent to $Z_{2i_0-3}Z_{2i_0-1}Z_{2i_0-1}X_{2i_0+1}=Z_{2i_0-3}X_{2i_0+1}$. Similarly, the local defect term $\mathcal{B}_{2i_0}^{\text{dTFI}}$ is equivalent to the three-site operator $X_{2i_0-1}Z_{2i_0+1}Z_{2i_0+3}$.

We are left with the matrix element for $\mathcal{C}_{2i_0}^{\text{dTFI}}$, namely $\langle h_{2i_0-1}\sigma h'_{2i_0+1}\sigma h'_{2i_0+3}\sigma |\mathcal{C}_{2i_0}^{\text{dTFI}}|h_{2i_0-1}\sigma h_{2i_0+1}\sigma h_{2i_0+3}\sigma \rangle$. This matrix element is represented diagrammatically in Figure \ref{fig:pauliZXX}. The second term on the RHS of Figure \ref{fig:pauliZXX} vanishes, using simplifications similar to those in Figure \ref{fig:pauliXX2}. Therefore we conclude that $\mathcal{C}_{2i_0}^{\text{dTFI}}$ is equivalent to the three-site operator $Z_{2i_0-1}X_{2i_0+1}X_{2i_0+3}$. 

We have reproduced the duality defect Hamiltonian, given in Eq.~(\ref{eqn:H_DdTFI}), using tools from the Ising fusion category. Essentially, local terms in the deformed model implemented by next-to-nearest-neighbor projectors are realized in a factorized fashion, using local terms coming from the nearest-neighbor projectors. This is expected to generalize to longer range projectors.

\insfigsvg{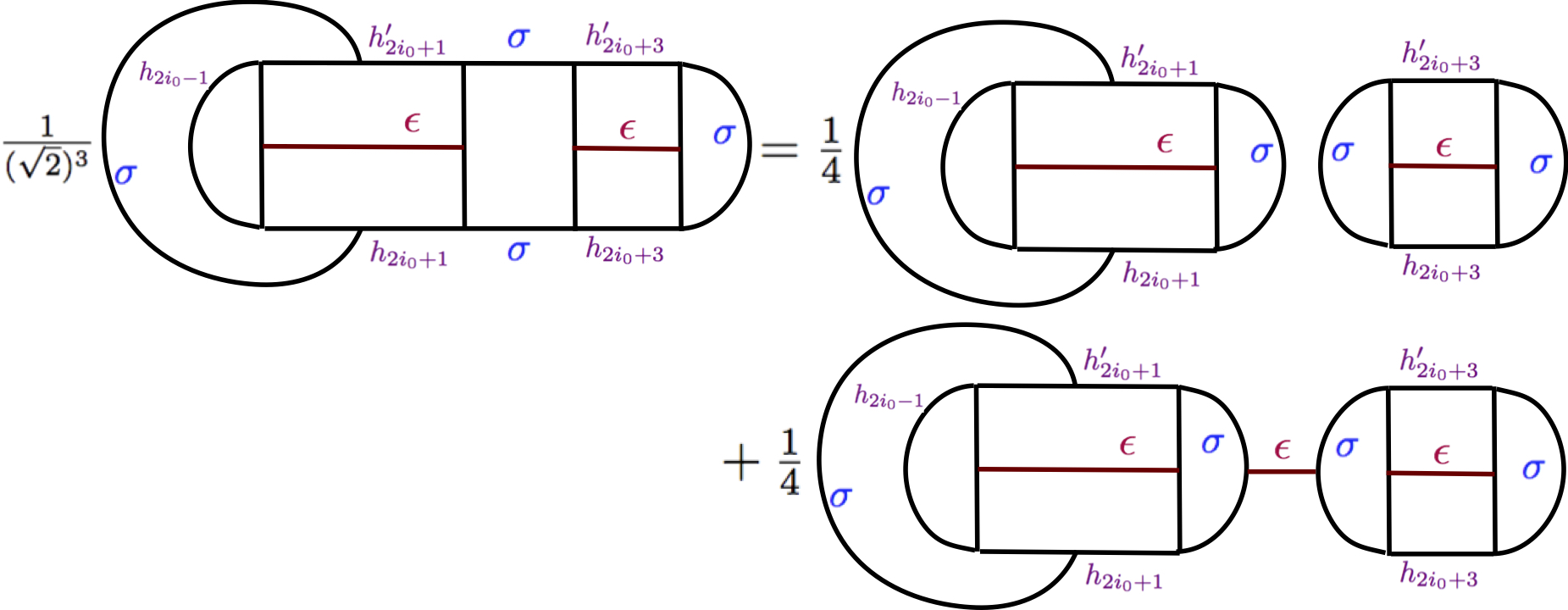}{0.14}{Diagrammatic presentation for the matrix element of $\mathcal{C}_{2i_0}^{\text{dTFI}}$.}

\section{Defect-modified weak integrability breaking }\label{sec:defectweakintebreaking}

Unlike generic (non-integrable) many-body systems, closed integrable systems don't thermalize as in the usual sense. Thermalization in integrable systems is described by the generalized Gibbs ensemble \cite{Kinoshita2006Nature,RigolPRL2007,Cassidy_2011,FagottiPRBXXZ,Essler:2016ufo}, which takes into account the additional conserved quantities apart from the total energy and the number of particles. In presence of an integrability breaking deformation, genuine thermalization is expected to happen again. For small deformations, the system first relaxes to a stationary state of the undeformed integrable Hamiltonian, while genuine thermalization happens at a later time \cite{Mitra_2018,Kollar_2011,Langen:2016vdb,Berges:2004ce}. For generic deformations, the thermalization time typically scales as $\lambda^{-2}$ where $\lambda$ is the deformation strength \cite{Mitra_2018,Mori:2017qhg,Mallayya:2021fdg}, as reasoned using Fermi's golden rule. However, for some specific systems the thermalization time can scale larger than $\lambda^{-2}$ \cite{Abanin_2017,Kurlov2022,Jung2006,Yeh:2023uyk,Yeh24a} and it takes longer for the system to thermalize. Recently, based on previous work on long-range deformations of integrable spin chains \cite{Bargheer:2008jt,Bargheer:2009xy,Pozsgay:2019ekd,Marchetto:2019yyt,Doyon:2021tzy}, the authors of \cite{Surace:2023wqq} described a systematic approach to engineer large families of deformations which break integrability more weakly than generic deformations, where the relaxation to thermal equilibrium upon introducing such weak integrability breaking deformations is slower than usual. In particular, the KW self-dual deformation in the TFI model is an instance of weak integrability breaking deformation.

In this Section, we will describe a method to construct defect-modified weak integrability breaking deformations, where the undeformed model is an integrable model with the insertion of an integrable defect. We will demonstrate this method by giving a third way to derive the duality defect Hamiltonian in the KW self-dual deformed TFI model, thanks to the fact that the KW duality defect in the undeformed TFI model is integrable. We also investigate additional defect-modified conserved charges (other than the Hamiltonian) in the TFI model, and confirm their slower decay behavior upon turning on the weak integrability breaking deformation. We will employ exact diagonalization for small system sizes to study the long time dynamics of infinite-temperature observables, where traditional tensor networks techniques are not well-suited. We refer to Ref. \cite{PhysRevB.110.075149} and Ref. \cite{PhysRevResearch.7.023245} for some recent developments of numerical methods to access such regimes of long time dynamics. Interestingly, as we discuss next, the physics of weak integrability breaking deformations is to cause a decay with a higher power of the integrability breaking term. This physics is apparent at short times itself, and therefore exact diagonalization is sufficient to demonstrate it.

\subsection{Weak integrability breaking deformations}\label{sec:weakintebreaking}

We first briefly review the construction of weak integrability breaking deformations following \cite{Surace:2023wqq}. Consider a 1d infinite lattice model with a set of {\it{extensive local}} conserved charges $Q_\alpha$, with
\begin{equation}
Q_\alpha = \sum\limits_j q_{\alpha,j}~,
\end{equation}
where $q_{\alpha,j}$ is the charge density operator with a finite support around the physical spin site $j$. The conserved charges, which in particular include the Hamiltonian, are mutually commuting. This model could be an integrable model, or a generic model with a finite set of conserved charges. 

Refs. \cite{Bargheer:2008jt,Bargheer:2009xy,Pozsgay:2019xak} considered smooth 1-parameter deformations of conserved charges, denoted as $Q_\alpha(\lambda)$, such that $\left[ Q_\alpha(\lambda) , Q_\beta(\lambda ) \right]=0$. This kind of deformations are generated by $X(\lambda)$ satisfying
\begin{equation}
\frac{d Q_\alpha(\lambda)}{d\lambda}=\text{i} \left[ X(\lambda), Q_\alpha(\lambda)\right] ~.
\end{equation}
The deformed charges $Q_\alpha(\lambda)$ should be extensive quasi-local \cite{Doyon:2021tzy}, namely the charge densities $q_{\alpha,j}(\lambda)$ should be supported around site $j$ with sufficiently decaying tails. This requirement puts constraints on the generator $X(\lambda)$. There are three large classes of $X(\lambda)$: extensive local or quasi-local operators, boosted operators, and bi-local operators. In particular, the KW self-dual deformation in the TFI model is generated by a bi-local operator, which we discuss in detail in Section \ref{sec:bilocalTFI}. 

The deformed Hamiltonian is a special instance of deformed charges. The $Q_\alpha(\lambda)$ and $H(\lambda)$ obtained in this method contain arbitrary long-range interactions whose strength decreases exponentially with the distance. To produce deformations with finite-range interactions, one expands $Q_\alpha(\lambda)$ and $H(\lambda)$ as a series in $\lambda$ and performs truncations, say up to order $l-1$, obtaining $Q_\alpha^{< l}(\lambda)$ and $H^{< l}(\lambda)$. The cost of doing so, is that $Q_\alpha^{< l}(\lambda)$ and $H^{< l}(\lambda)$ don't strictly commute anymore. Rather, they commute up to a correction at order $\lambda^l$. Concretely we have
\begin{equation}
\begin{split}
\left[  Q_\alpha^{<l}(\lambda),Q_{\beta}^{<l}(\lambda) \right] & = o(\lambda^l) ~ ,\\
\left[  Q_\alpha^{<l}(\lambda),H^{<l}(\lambda) \right] & = o(\lambda^l) ~ .\\
\end{split}
\end{equation}
The special case most-relevant to us will be the case of $l=2$, namely
\begin{equation}
\begin{split}
Q_\alpha^{<2}(\lambda)&=Q_\alpha^{(0)}+\lambda Q_\alpha^{(1)} ~ ,\\
H^{<2}(\lambda) &= H^{(0)}+\lambda H^{(1)} ~ .
\end{split}
\end{equation}
The first-order corrections $Q_\alpha^{(1)}$ (including $H^{(1)}$) are entirely determined by the undeformed conserved charges $Q_\alpha^{(0)}$ (including the undeformed Hamiltonian) and the $0^{\text{th}}$-order generator $X^{(0)}$:
\begin{equation}\label{eqn:1storder}
\begin{split}
Q_\alpha^{(1)}&=\text{i}\left[ X^{(0)}, Q_\alpha^{(0)} \right]~,\\
H^{(1)}&=\text{i}\left[X^{(0)},H^{(0)} \right]~.
\end{split}
\end{equation}

For a system described by the deformed Hamiltonian $H^{<2}(\lambda)$, there exists a family of quasi-conserved charges $Q_\alpha^{<2}(\lambda)$, which almost mutually commute with each other up to corrections at $o(\lambda^2)$. In other words, the effective integrability breaking strength is quadratic in the deformation strength. In this sense, this kind of deformations break integrability more weakly than a generic deformation.

\subsection{Self-dual deformation in the TFI model}\label{sec:bilocalTFI}

In this section, we will put the descriptions in Section \ref{sec:weakintebreaking} in the concrete context of the self-dual deformed TFI model, and consider the decay of deformed charges. The construction of this deformation involves 2 commuting charges in the undeformed TFI model \cite{Surace:2023wqq}: 
\begin{equation}\label{eqn:q1q2nodefect}
\begin{split}
Q_1^{(0)}&=\sum\limits_j q^{(0)}_{1,j}=\sum\limits_j (-Y_{j} Z_{j+1}+Z_{j} Y_{j+1}) ,\\
Q_2^{(0)}=-\frac{1}{J}H^{(0)}&=\sum\limits_j q^{(0)}_{2,j}=\sum\limits_j Z_j Z_{j+1} +\frac{h}{2} (X_j + X_{j+1}) 
\end{split}
\end{equation}
where $h=g/J$ for the convention that we use in 
Eq.~(\ref{eqn:HdTFI}) and we again consider an infinite chain.

The KW self-dual deformation in the TFI model is generated by a bi-local operator constructed using $Q_1^{(0)}$ and $Q_2^{(0)}$. In general, given two extensive local operators $\mathcal{O}_\alpha$ and $\mathcal{O}_\beta$, one can define the following bi-local operator \cite{Bargheer:2009xy,Pozsgay:2019xak}, which roughly speaking can be viewed as ``one half" of the anti-commutator with a choice of ordering
\begin{equation}\label{eqn:bilocaldef}
\left[\mathcal{O}_\alpha\big|\mathcal{O}_\beta\right]=\sum\limits_{j<k}\{o_{\alpha,j},o_{\beta,k} \} + \frac{1}{2}\sum\limits_{j}\{ o_{\alpha, j}, o_{\beta,j}  \}~,
\end{equation}
where $o_{\alpha,j}$ and $o_{\beta,k}$ are the corresponding local on-site density operators. For mutually commuting conserved charges $Q_\alpha^{(0)}$ with local densities $q_{\alpha,j}$, it is more convenient to do computations through generalized current densities, defined through
\begin{equation}
\text{i}\left[ Q_\alpha^{(0)}, q^{(0)}_{\beta,j} \right] = J^{(0)}_{\beta\alpha,j}-J^{(0)}_{\beta\alpha,j+1}~.
\end{equation}

Notice that, in the special case where $Q_\alpha^{(0)}$ is the Hamiltonian, one would obtain the ordinary current densities, as the LHS would describe the time derivative of the charge density.

To compute the first order deformations $Q_\alpha^{(1)}$, we then use Eq.~(\ref{eqn:1storder}) with $X^{(0)}=\left[Q_2^{(0)}\big| Q_1^{(0)}\right]$. In particular, $Q_\alpha^{(1)}$ can be expressed in terms of generalized current densitites as
\begin{equation}\label{eqn:bilocaldefcurrent}
\begin{split}
Q_{\alpha}^{(1)} &= \frac{1}{2}\sum\limits_j\{q^{(0)}_{1,j}, J^{(0)}_{2\alpha,j}+J^{(0)}_{2\alpha,j+1}  \}\\
&-\frac{1}{2}\sum\limits_j\{q^{(0)}_{2,j}, J^{(0)}_{1\alpha,j}+J^{(0)}_{1\alpha,j+1}  \} ~,
\end{split}
\end{equation}
where the charge densities $q^{(0)}_{1,j}$ and $q^{(0)}_{2,j}$ are given as in Eq. (\ref{eqn:q1q2nodefect}).

We can now apply Eq. \eqref{eqn:bilocaldefcurrent} to compute first-order deformations. The computational details can be found in Appendix \ref{sec:appweakintebreaking}. Concretely the first-order deformation to the Hamiltonian, $H^{(1)}=-JQ_2^{(1)}$, is given by
\begin{equation}\label{eqn:H1nodefect}
H^{(1)} = -4 g \left(\sum\limits_j Z_{j}Z_{j+2} + \sum\limits_j X_j X_{j+1} \right).
\end{equation}
Rescaling the deformation parameter $\lambda$ by a factor of $4g$, we see that $H^{<2}(\lambda)=H^{(0)}+\lambda H^{(1)}$ indeed reproduces the Hamiltonian for the KW self-dual deformed TFI model. We also remark that, to obtain this result, we have relabeled the summation indices for certain terms, which is justified for the infinite chain setup (as well as for a periodic chain).

We now proceed to compute the deformed charge $Q_1^{<2}(\lambda)=Q_1^{(0)}+\lambda Q_1^{(1)}$, where the undeformed charge $Q_1^{(0)}$ is given in Eq. (\ref{eqn:q1q2nodefect}). We obtain
\begin{equation}\label{eqn:Q1nodefect}
\begin{split}
Q_1^{(1)}=&\frac{h}{g}\left(- \sum\limits_j Y_j Z_{j+2}+\sum\limits_j Z_j Y_{j+2} \right) \\
&+ \frac{1}{g}\left( \sum\limits_j Z_jY_{j+1} X_{j+2} - \sum\limits_j X_j Y_{j+1}Z_{j+2} \right),
\end{split}
\end{equation}
with $h=g/J$ in the convention of Eq. (\ref{eqn:HdTFI}).

Up until now, we have been working with an infinite chain. In fact, from Eq. (\ref{eqn:bilocaldef}), it is clear that the bi-local generator $X^{(0)}=\left[Q_2^{(0)}\big|Q_1^{(0)}\right]$ requires a choice of ordering, which raises potential issues for a finite chain with e.g. periodic boundary condition. On the other hand, the first order deformations $Q_\alpha^{(1)}=\text{i}\left[\left[Q_2^{(0)}\big|Q_1^{(0)}\right],Q_\alpha^{(0)}\right]$ (as in Eq. (\ref{eqn:bilocaldefcurrent})) can be expressed in terms of local charge density operators and generalized current density operators, which readily carry over to the case of a periodic chain. In particular, we can apply the resulting deformations such as $H^{(1)}$ and $Q_1^{(1)}$ to our original setup of a periodic chain with $L$ sites. Moreover, we explicitly checked that
\begin{equation}
\left[ H^{(1)}, Q_1^{(0)} \right]+\left[ H^{(0)}, Q_1^{(1)} \right]=0~
\end{equation}
for the periodic chain setup, which implies that indeed 
\begin{equation}
\left[H^{<2}(\lambda),Q_1^{<2}(\lambda)\right]=o(\lambda^2)~.
\end{equation}

We now proceed to investigate the decay behavior of the deformed charge $Q_1^{<2}(\lambda)$, under the time evolution governed by the KW self-dual deformed Hamiltonian in Eq. (\ref{eqn:HdTFI}). Because $Q_1^{<2}(\lambda)$ commutes with the deformed Hamiltonian with corrections of order $\lambda^2$, as argued in \cite{Surace:2023wqq}, one would expect slower decay of the deformed charge compared with an arbitrary deformation.

Our observable here is the infinite-temperature auto-correlation function of $Q_1^{<2}(\lambda)$, defined as
\begin{equation}\label{eqn:AinftydefQ}
A_\infty^{Q_1^{<2}(\lambda)}(t):=\frac{1}{\mathcal{N}}\text{Tr}\left(\text{e}^{\text{i}H^{<2}(\lambda)t}Q_1^{<2}(\lambda)\text{e}^{-\text{i}H^{<2}(\lambda)t}Q_1^{<2}(\lambda)\right)
\end{equation}
where the normalization $\mathcal{N}$ is given by
\begin{equation}
\mathcal{N}=\text{Tr}\left(Q_1^{<2}(\lambda)Q_1^{<2}(\lambda)\right)~,
\end{equation}
such that $A_\infty^{Q_1^{<2}(\lambda)}(t=0)=1$. 

We numerically study $A_\infty^{Q_1^{<2}(\lambda)}(t)$ for small system sizes through Exact Diagonalization. After an $o(1)$ initial decay time, $A_\infty^{Q_1^{<2}(\lambda)}(t)$ oscillates around a plateau. This is demonstrated in the left panel of Figure \ref{fig:DecayNoDefect}. Moreover, similar to \cite{Surace:2023wqq}, we can perform a perturbative analysis for the decay behavior at early times. The details can be found in Appendix \ref{sec:decayrate}. In particular, the $o(\lambda^2)$ dependence of $\frac{d}{dt}A_\infty^{Q_1^{<2}(\lambda)}(t)$ is expected to vanish, such that its leading-order dependence is at least $o(\lambda^3)$. In the right panel of Figure \ref{fig:DecayNoDefect}, we plot $A_\infty^{Q_1^{<2}(\lambda)}(t)$ against $\lambda^3 t$ for different system sizes and different values of $\lambda$ and observe good data points collapse at early times.  

\newpage

\onecolumngrid

\insfigsvg{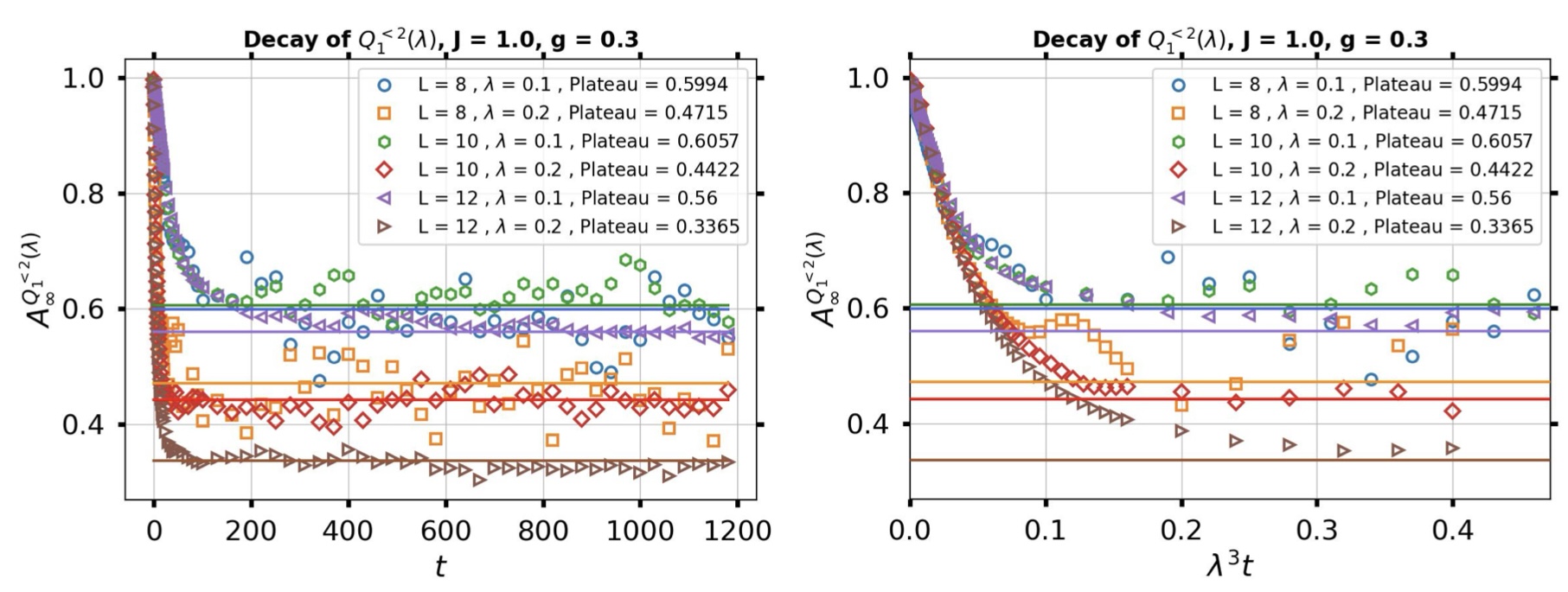}{0.25}{{\it Left Panel} : The infinite-temperature auto-correlation function $A_\infty^{Q_1^{<2}(\lambda)}$ as defined in Eq. \eqref{eqn:AinftydefQ}, plotted against time $t$ for $J=1$, $g=0.3$, $L=8,10,12$ and $\lambda=0.1,0.2$. {\it Right Panel} : Short-time decay behavior of $A_\infty^{Q_1^{<2}(\lambda)}$ plotted against $\lambda^3 t$ for the same set of parameters. The data points collapse at early times indicates an $o(\lambda^3)$ leading-order dependence.}

\twocolumngrid

\subsection{Weak integrability breaking in presence of KW duality defect}\label{sec:weakintegrabilitybreakingKW}

In this section, we will describe how the above story of weak integrability breaking deformations can be generalized to cases with the insertion of defects. In integrable models, the insertion of an integrable defect $D$ can be realized by local modification of the transfer matrix without spoiling integrability. In particular, one can obtain the defect Hamiltonian from the locally-modified transfer matrix as usual. More generally though, additional higher conserved charges $Q_\alpha^{(0)}$ are also modified by the defect, and we denote the modified conserved charges as $Q_\alpha^{D,(0)}$. We can now play the game of weak integrability breaking deformation using $Q_\alpha^{D,(0)}$. In particular, suppose that for the bulk theory there exists a certain weak integrability breaking deformation generated by e.g. a bi-local operator $\left[Q_\alpha^{(0)}\big|Q_\beta^{(0)}\right]$. Then, for the same bulk theory but now with the insertion of an integrable defect $D$, the bi-local operator $\left[ Q_\alpha^{D,(0)}\big|Q_\beta^{D,(0)}\right]$ generates the corresponding weak integrability breaking deformation in the defect-modified theory. In particular, one should be able to derive the defect Hamiltionian in the deformed theory in this way.

The KW duality defect in the TFI model is an instance of an integrable defect. In the following, we will showcase defect-modified weak integrability breaking using this example. In particular, this gives us a third way to derive the defect Hamiltonian (given by $-JQ_2^{D,<2}(\lambda)$) in the deformed TFI model. Additionally, we study the decay of deformed charges in presence of the defect.

We remark that, in the main text of this manuscript, we have been mostly considering the situation where the defect is realized as a KW duality interface composed with a domain wall in the coupling space. For the demonstration of the weak integrability breaking method though, we will consider the insertion of a KW duality interface alone on an infinite chain. Once the deformed charges $Q_\alpha^{D,<2}$ are computed, composing with a domain wall could bring us back to the situation considered in the main text.
 
Our first step would be to find the modification to the charges $Q_1^{(0)}$ and $Q_2^{(0)}$ in Eq. (\ref{eqn:q1q2nodefect}). Recall from Section \ref{sec:halfKW} that, a duality interface can be constructed by performing a half-chain KW transformation. Concretely, the KW transformation (composed with a half-site translation) acts locally on the Pauli operators that appear in the Hamiltonian of the TFI model in the following way:
\begin{equation}
\widehat{D} X_j =Z_j Z_{j+1}\widehat{D}, ~ \widehat{D} Z_j Z_{j+1} = X_{j+1}\widehat{D}~.
\end{equation}
When one performs the KW transformation to the right of, say, site $i_0$, away from this interface the KW transformation (denoted as $\widehat{D}_{i_0}$) acts as before. Right at the interface we have 
\begin{equation}
\widehat{D}_{i_0} Z_{i_0}Z_{i_0+1}= Z_{i_0}X_{i_0+1} \widehat{D}_{i_0} ~ .
\end{equation}
Performing the half-chain KW transformation then yields the following duality interface Hamiltonian density:
\begin{equation}
\begin{split}
q^{D,(0)}_{2,j<i_0}&=Z_jZ_{j+1}+\frac{h}{2}X_j+\frac{h}{2}X_{j+1}~,\\
q^{D,(0)}_{2,i_0} &= Z_{i_0} X_{i_0+1} +\frac{h}{2} X_{i_0}~,\\
q^{D,(0)}_{2,i_0+1}&= hZ_{i_0+1}Z_{i_0+2}+\frac{1}{2} X_{i_0+2}~,\\
q^{D,(0)}_{2,j>i_0+1}&=hZ_jZ_{j+1}+\frac{1}{2}X_j+\frac{1}{2}X_{j+1}~.
\end{split}
\end{equation}

We can similarly work out the modification to $Q_1^{(0)}$. First of all, notice that $Q_1^{(0)}$ is self-dual under KW transformation, in particular we have
\begin{equation}
\widehat{D} Y_j Z_{j+1} = - Z_j Y_{j+1} \widehat{D}, ~ \widehat{D} Z_j Y_{j+1}=-Y_{j+1}Z_{j+2}\widehat{D}~.
\end{equation}
Performing the half-space KW transformation, at the interface we have
\begin{equation}
\begin{split}
\widehat{D}_{i_0}Y_{i_0}Z_{i_0+1}& = Y_{i_0} X_{i_0+1}\widehat{D}_{i_0} ~ ,\\
\widehat{D}_{i_0}Z_{i_0}Y_{i_0+1} & = -\text{i}Z_{i_0}X_{i_0+1}Z_{i_0+1}Z_{i_0+2}\widehat{D}_{i_0}\\
&=-Z_{i_0}Y_{i_0+1}Z_{i_0+2}\widehat{D}_{i_0} ~ .
\end{split}
\end{equation}
This leads to the following modification to the charge density originally given in Eq. (\ref{eqn:q1q2nodefect}):
\begin{equation}\label{eqn:q1defect}
\begin{split}
q_{1,i_0}^{D,(0)}&= -Y_{i_0}X_{i_0+1}-Z_{i_0}Y_{i_0+1}Z_{i_0+2} ~ ,\\
q_{1,i_0+1}^{D,(0)}&=Z_{i_0+1}Y_{i_0+2} ~.
\end{split}
\end{equation}
As a sanity check, we can show that
\begin{equation}
\left[ Q_2^{D,(0)} , Q_1^{D,(0)} \right]=0 ~,
\end{equation}
where the details can be found in Appendix \ref{sec:appweakintebreaking}.

Now we are ready to study the deformation generated by the defect-modified bilocal operator $\left[ Q_2^{D,(0)}\big| Q_1^{D,(0)} \right]$. Concretely, the first order deformation to the defect Hamiltonian in integrable defect TFI model is 
\begin{equation}
\begin{split}
H^{D,(1)}&=-\frac{J}{2}\sum\limits_j \{q_{1,j}^{D,(0)}, J_{22,j}^{D,(0)}+J_{22,j+1}^{D,(0)} \}\\
&+\frac{J}{2}\sum\limits_j \{q_{2,j}^{D,(0)}, J_{12,j}^{D,(0)}+J_{12,j+1}^{D,(0)} \}\\
=&-4g\bigg(\sum\limits_{\substack{j\neq i_0-1,\\j\neq i_0}}Z_j Z_{j+2}+\sum\limits_{\substack{j\neq i_0,\\j\neq i_0+1}}X_j X_{j+1}\\
&+Z_{i_0-1}X_{i_0+1}+X_{i_0}Z_{i_0+1}Z_{i_0+2}+Z_{i_0}X_{i_0+1}X_{i_0+2}\bigg).
\end{split}
\end{equation}
Rescaling $\lambda \to 4g\lambda$ as before, this reproduces the deformation terms, including the extra local modifications, in the defect Hamiltonian Eq. (\ref{eqn:H_DdTFI1}) for the self-dual deformed duality-twisted TFI model.

Next, we will compute the deformed charge $Q_1^{D,<2}(\lambda)$ and investigate its decay under the deformed defect Hamiltonian $H^{D,<2}(\lambda)$. From Eq. (\ref{eqn:q1q2nodefect}) and Eq. (\ref{eqn:q1defect}), the undeformed duality-twisted charge is given by
\begin{equation}
\begin{split}
Q_1^{D,(0)}&=\sum\limits_{\substack{j,\\ j\neq i_0,i_0+1}} \left(-Y_jZ_{j+1}+Z_j Y_{j+1}\right)\\
&-Y_{i_0}X_{i_0+1}-Z_{i_0}Y_{i_0+1}Z_{i_0+2}+Z_{i_0+1}Y_{i_0+2}~.
\end{split}
\end{equation}
The corresponding $o(\lambda)$ deformation $Q_1^{D,(1)}$ is given by (see Appendix \ref{sec:appweakintebreaking})
{\small{\begin{equation}\label{eqn:Q11defect}
\begin{split}
&Q_1^{D,(1)}=\frac{1}{J}\sum\limits_{j\leq i_0-2} (Z_jY_{j+2}-Y_jZ_{j+2})\\
&+\frac{1}{g}\sum\limits_{j\leq i_0-2} (Z_jY_{j+1}X_{j+2}-X_jY_{j+1}Z_{j+2})\\
&+\frac{1}{g}\sum\limits_{j\geq i_0+2} (Z_jY_{j+2}-Y_jZ_{j+2})\\
&+\frac{1}{J}\sum\limits_{j\geq i_0+2} (Z_jY_{j+1}X_{j+2}-X_jY_{j+1}Z_{j+2})\\
&+\frac{1}{g}Z_{i_0+1}Y_{i_0+3}+\frac{1}{J}Z_{i_0+1}Y_{i_0+2}X_{i_0+3}-\frac{1}{J}Y_{i_0-1}X_{i_0+1}\\
&-\frac{1}{g}X_{i_0-1}Y_{i_0}X_{i_0+1}+\frac{1}{g}X_{i_0}Z_{i_0+1}Y_{i_0+2}-\frac{1}{g}Z_{i_0}Y_{i_0+1}Z_{i_0+3}\\
&-\frac{1}{J} Z_{i_0-1}Y_{i_0+1}Z_{i_0+2}-\frac{1}{J}Y_{i_0}X_{i_0+1}X_{i_0+2}\\
&+\frac{1}{g}Z_{i_0-1}Y_{i_0}Z_{i_0+1}Z_{i_0+2}-\frac{1}{J}Z_{i_0}X_{i_0+1}Y_{i_0+2}Z_{i_0+3}
\end{split}
\end{equation}}}
where we have again applied the rescaling convention of $\lambda\to 4g\lambda$. Notice that, $Q_1^{D,(1)}$ can also be constructed from performing half-space KW transformation on $Q_1^{(1)}$ in the no-defect case. We have checked that the results do agree from these two methods.

Similar to Section \ref{sec:bilocalTFI}, we proceed to study the decay of the defect modified deformed charge $Q_1^{D,<2}(\lambda)$, where our observable is the infinite-temperature auto-correlation function $A_\infty^{Q_1^{D,<2}(\lambda)}(t)$ defined in the same way as in Eq. (\ref{eqn:AinftydefQ}). We numerically study this quantity at the critical point where $J=g$. After an initial decay time, $A_\infty^{Q_1^{D,<2}(\lambda)}(t)$ oscillates around a plateau. Moreover, similar to the no-defect case, the leading-order dependence of the early time decay behavior is expected to be $o(\lambda^3)$. This is supported by the data points collapse at early times as illustrated in Figure \ref{fig:DecayCollapseDefect}.

\insfigsvg{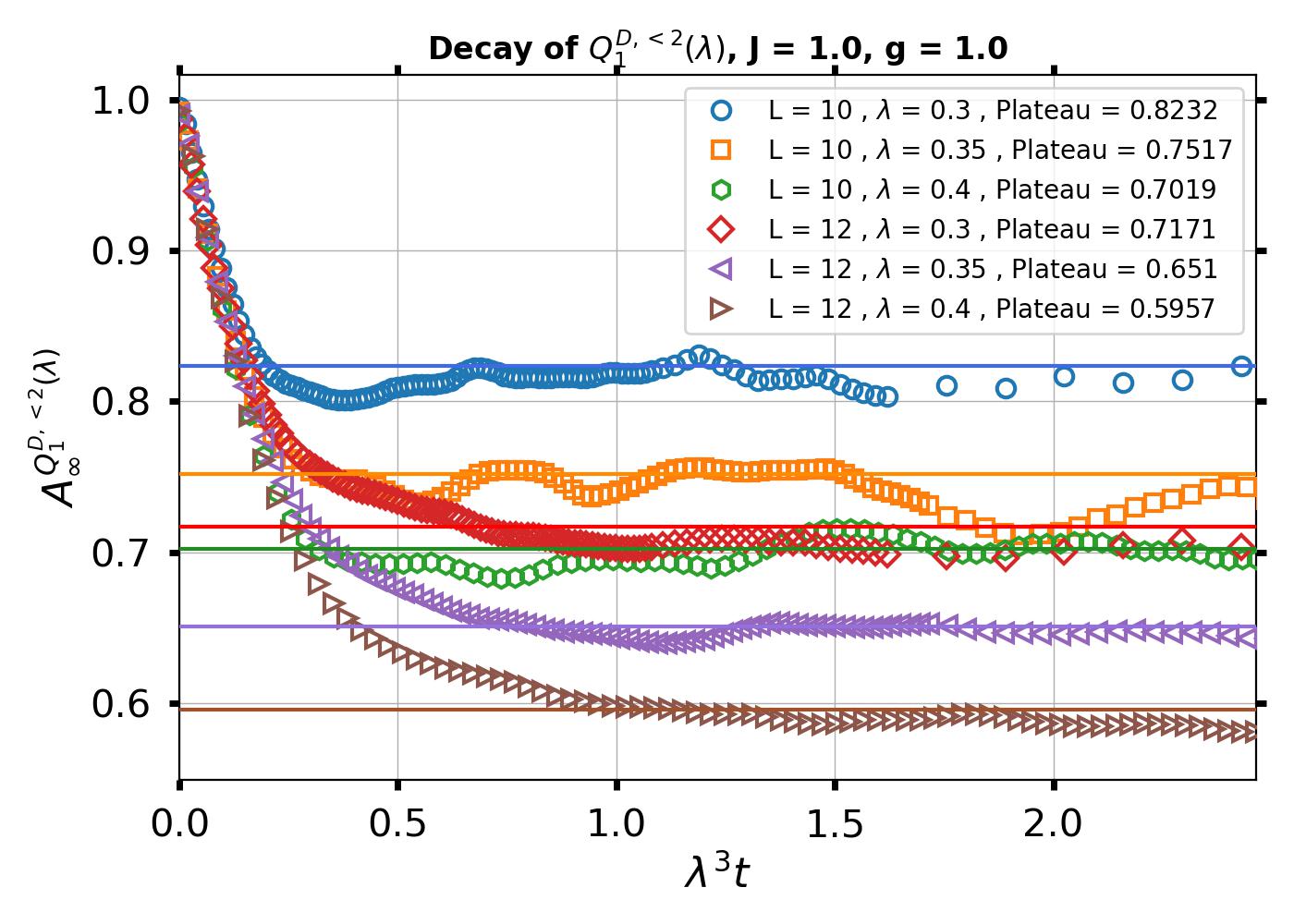}{0.5}{Short-time decay behavior of $A_\infty^{Q_1^{D,<2}(\lambda)}(t)$ for $J=1$, $g=1$, $L=10,12$ and $\lambda=0.3, 0.35, 0.4$ respectively. The data points collapse at early times indicates an $o(\lambda^3)$ leading-order dependence.}

We conclude this Section with some comments. It is known that, for the off-critical TFI model on a circular chain with a duality interface and a domain wall, there exists a Majorana zero mode localized at the domain wall separating the two gapped phases \cite{Fendley16}. In the setup where the duality interface is right next to the domain wall, this zero mode becomes localized at the duality defect. A natural question is whether one can construct a deformation of the zero mode using the method described above. However, we can quickly check that, although the zero mode (by construction) commutes with the defect Hamiltonian in the integrable TFI model, it doesn't commute with the higher conserved charge $Q_1^{D,(0)}$. Therefore the weak integrability breaking deformation generated by the bi-local operator $\left[Q_2^{D,(0)}\big|Q_1^{D,(0)}\right]$ does not apply for the Majorana zero mode.

\section{Duality defect in the deformed Floquet TFI model}\label{sec:FloquetdTFIdefect}

We now turn to the periodically driven Floquet TFI model. First, consider the undeformed Floquet TFI model on a periodic chain of $L$ sites. The Floquet unitary governing the time evolution within a driving period is given by  
\begin{equation}\label{eqn:FloquetTFI}
U_{\text{TFI}} = \left(\prod\limits_{j=1}^L \text{e}^{-\text{i}u_j^b  X_j}\right)\left(\prod\limits_{j=1}^L \text{e}^{-\text{i}u_j^a  Z_j Z_{j+1}}\right)~,
\end{equation}
where the driving period has been absorbed into the couplings $u_j^{a,b}$, and we have specified a particular Trotter choice. In general, the couplings $u_{j}^{a,b}$ can be made arbitrary. In the context of this manuscript, we will restrict to uniform couplings $u_j^a=JT/2$ and $u_j^b=gT/2$.  

In the Floquet TFI model described by Eq. (\ref{eqn:FloquetTFI}), one can introduce a KW duality defect, as studied in \cite{Tan22}. Concretely, the Floquet unitary is locally modified between sites $i_0$ and $i_0+1$ as
\begin{equation}\label{eqn:FloquetTFIdefect}
\begin{split}
U_{D,\text{TFI}}=&\left(\prod\limits_{\substack{j=1,\\ j\neq i_0+1}}^L \text{e}^{-\text{i}u_j^b X_j}\right)\times {\color{purple}\text{e}^{-\text{i}u^a_{i_0}Z_{i_0}X_{i_0+1}}}\\
&\times \left(\prod\limits_{\substack{j=1,\\j\neq i_0}}^L \text{e}^{-\text{i}u_j^a  Z_j Z_{j+1}}\right)~.
\end{split}
\end{equation}

To solve this model, we perform the Jordan-Wigner transformation. We adopt the convention that the string operators are attached to site $i_0+1$ (where the transverse field is missing) and wind to the right of the defect. The Majorana fermion operators are defined as
\begin{equation}
\begin{split}
\gamma_{2j}&=\left(\prod\limits_{k=i_0+1}^{j-1} X_k\right) Z_j~, j\neq i_0+1\\
\gamma_{2j+1}&=-\left(\prod\limits_{k=i_0+1}^{j-1} X_k\right) Y_j~, j\neq i_0+1\\
\gamma_{2i_0+2}&=Z_{i_0+1}~,~ \gamma_{2i_0+3}=-Y_{i_0+1}~.
\end{split}
\end{equation}
In terms of the Majorana fermion operators, the $\mathbb{Z}_2$ charge $\Omega$ in Eq. (\ref{eqn:Z2QTFIdefect}) is given as
\begin{equation}
\Omega =(-\text{i})^{L-1} \gamma_2\dots\gamma_{2i_0+1}\gamma_{2i_0+3}\dots\gamma_{2L}\gamma_{2L+1}~.
\end{equation}
The Floquet unitary in Eq. (\ref{eqn:FloquetTFIdefect}) becomes
\begin{equation}\label{eqn:fermionFloquetTFIdefect}
\begin{split}
U_{D,\text{TFI}}=&\left(\prod\limits_{\substack{j=1,\\ j\neq i_0+1}}^L \text{e}^{- u_j^b  \gamma_{2j}\gamma_{2j+1}}\right)\times \text{e}^{-u^a_{i_0}\Omega \gamma_{2i_0+1}\gamma_{2i_0+3}}\\
&\times \left(\prod\limits_{\substack{j=1,\\j\neq i_0}}^L \text{e}^{- u_j^a \gamma_{2j+1}\gamma_{2j+2}}\right)~. 
\end{split}
\end{equation}
Notice that $\gamma_{2i_0+2}$ is absent from $U_{D,\text{TFI}}$, however it doesn't commute with $U_{D,\text{TFI}}$ due to $\{\gamma_{2i_0+2},\Omega\}=0$. Within each $\mathbb{Z}_2$ charge sector labeled by $\Omega=\pm 1$, the Floquet unitary $U_{D,\text{TFI}}$ is quadratic in Majorana fermions and the dynamics is exactly solvable. Moreover, for fixed $\Omega =\pm 1$, the dynamics of $U_{D,\text{TFI}}$ only involves an odd number of Majorana fermions. By a counting argument, there must exist at least one Majorana zero mode $\Psi_0$ within each $\Omega=\pm 1$ sector \cite{Tan22}, satisfying $\left[ \Psi_0, U_{D,\text{TFI}}\right]=0$. In the case of uniform couplings $u_j^a=JT/2$ and $u_j^b=gT/2$, it was shown in \cite{Tan22} that $\Psi_0$ is localized around the duality defect, with a width of
\begin{equation}
\xi \sim \left(\text{ln}\frac{\text{tan}\left(\text{max}(J,g)\right)}{\text{tan}\left(\text{min}(J,g)\right)}\right)^{-1}~.
\end{equation}
In the special case of $g=0$, we have \cite{Tan22}
{\small{\begin{equation}\label{eqn:zeromodeg0}
\begin{split}
\Psi_0&=\frac{1}{\sqrt{1+\text{sec}^2\frac{JT}2}}\left( \Omega \gamma_{2i_0+1} +\text{tan}\frac{JT}{2}\gamma_{2i_0+3}+\gamma_{2i_0+4}\right)\\
&=\frac{1}{\sqrt{1+\text{sec}^2\frac{JT}2}}\left( Z_{i_0}Z_{i_0+1} -\text{tan}\frac{JT}{2}Y_{i_0+1}+X_{i_0+1}Z_{i_0+2}\right).
\end{split}
\end{equation}}}

Turning on $\lambda\neq 0$ corresponds to turning on a four-fermion interaction on the Majorana fermion side. In the following, we will investigate the fate of the Majorana zero mode in the duality defect twisted deformed TFI model.  

As a first step, we fix our Trotter convention for the Floquet unitary in the deformed TFI model with periodic boundary condition:
\begin{equation}\label{eqn:dFloquetTFI}
\begin{split}
U_{\text{dTFI}} &=  \left(\prod\limits_{j=1}^{L}\text{e}^{-\text{i}v_j^b X_jX_{j+1}} \right)\left(\prod\limits_{j=1}^{L}\text{e}^{-\text{i}v_j^a Z_jZ_{j+2}} \right)\\
&\times \left(\prod\limits_{j=1}^{L}\text{e}^{-\text{i}u_j^b X_j} \right)\left(\prod\limits_{j=1}^{L}\text{e}^{-\text{i}u_j^a Z_jZ_{j+1}} \right)~.
\end{split}
\end{equation}
In the generic setup of Floquet circuit, one can take the couplings $u_j^{a,b},v_j^{a,b}$ to be non-uniform in space. In the context of this work, we again take the following uniform couplings
\begin{equation}\label{eqn:uniformcouplings}
u_{j}^a=J\frac{T}{4}, ~ u_j^b=g\frac{T}{4}, ~ v_j^a=v_j^b=\lambda \frac{T}{4}~.
\end{equation}

Now we introduce a duality defect which modifies the deformed TFI Floquet unitary locally around certain sites. An important step is to determine the relative ordering of the non-commuting terms in the Floquet unitary. In particular, given a Trotter choice for the Floquet unitary in the no-defect model, we need a prescription for the ordering of non-commuting terms in the Floquet unitary with the insertion of a duality defect. In \cite{Tan22}, this prescription was given in the language of face models of \cite{Fendley16}. In order to generalize such prescription to the deformed Floquet TFI model, we first reinterpret this prescription in terms of a half-chain sequential KW transformation, then apply it to obtain the Floquet unitary for the duality defect twisted deformed TFI model. The detailed ordering prescription and derivation for the Floquet unitary can be found in Appendix \ref{sec:FloOrdering}. 

In the end, we arrive at the following Floquet unitary for the deformed TFI model with the insertion of a KW duality defect:
{\small{\begin{equation}\label{eqn:dFloquetDTFI}
\begin{split}
U_{D,\text{dTFI}}&=\left(\prod\limits_{\substack{j=1,\\ j\neq i_0,\\i_0+1,\\i_0+2}}^L\text{e}^{-\text{i}v_j^b X_j X_{j+1}} \right) {\color{purple}\text{e}^{-\text{i}v_{i_0}^b X_{i_0}Z_{i_0+1}Z_{i_0+2}}}\\
\times& \text{e}^{-\text{i}v_{i_0+2}^b X_{i_0+2}X_{i_0+3}}{\color{purple}\text{e}^{-\text{i}v_{i_0-1}^a Z_{i_0-1}X_{i_0+1}}}\\
\times& {\color{purple}\text{e}^{-\text{i}v_{i_0}^a Z_{i_0}X_{i_0+1}X_{i_0+2}}}\left(\prod\limits_{\substack{j=1,\\ j\neq i_0-1,\\
i_0}}^L\text{e}^{-\text{i}v_j^a Z_j Z_{j+2}}\right)U_{D,\text{TFI}}, 
\end{split}
\end{equation}}}
where $U_{D,\text{TFI}}$ is the duality defect twisted Floquet unitary (Eq. (\ref{eqn:FloquetTFIdefect})) in the undeformed TFI model, and the couplings are taken to be uniform in space as given by Eq. (\ref{eqn:uniformcouplings}).

With the $\lambda\neq 0$ deformation turned on, the zero mode $\Psi_0$ in the undeformed Floquet TFI model is not preserved anymore and is expected to decay. We can numerically study its decay via the infinite-temperature auto-correlation function of a certain local operator which has a non-trivial overlap with $\Psi_0$. An optimal choice is the operator $Y_{i_0+1}$, as suggested by Eq. (\ref{eqn:zeromodeg0}). Namely, our observable is 
{\small{\begin{equation}\label{eqn:AinftyY}
A_\infty^{Y_{i_0+1}}(\lambda,n):= \frac{1}{2^L} \text{Tr}\left[ \left(U_{D,\text{dTFI}}^{\dagger}\right)^n Y_{i_0+1} \left(U_{D,\text{dTFI}}\right)^n Y_{i_0+1} \right]~,
\end{equation}}}
where $n$ is the number of time steps in the Floquet system.

Numerical analysis reveals that, for relatively small-sized systems, the zero mode $\Psi_0$ only partially decays. This is corroborated by the behavior of the infinite-temperature auto-correlation function $A_\infty^{Y_{i_0+1}}(\lambda,n)$. After an initial decay, $A_\infty^{Y_{i_0+1}}(\lambda,n)$ reaches a system-size dependent plateau given by 
\begin{equation}\label{eqn:plateauheight}
P_{A_\infty^{Y_{i_0+1}}}(\lambda,L)=\frac{1}{2^L}\sum\limits_{\beta}|\langle \beta| Y_{i_0+1}|\beta\rangle|^2~,
\end{equation}
where $|\beta\rangle$ are the eigenstates of the Floquet unitary $U_{D,\text{dTFI}}$. As some examples, in the left panel of Figure \ref{fig:FloquetAYDecayDefect} we plot $A_\infty^{Y_{i_0+1}}(\lambda,n)$ for $J=1$, $g=0.3$, $T=4$ with varying deformation strength $\lambda$ and system size $L$. The numerical plateau heights agree well with the values given by Eq. ($\ref{eqn:plateauheight}$). Moreover, the plateau height decreases as the system size increases. Further extrapolation indicates that, as $L\to \infty$, the plateau height approaches zero, indicating its nature of a finite-size effect.

We now take a closer look at the two stages of $A_\infty^{Y_{i_0+1}}(\lambda,n)$: the initial decay stage and the plateau stage. 
For the parameter choice of $J=1$, $g=0.3$ and $T=4$, numerical analysis suggests that the initial decay of $A_\infty^{Y_{i_0+1}}$ has a leading $o(\lambda^3)$ dependence on the deformation strength, as indicated by the data points collapse shown in the right panel of Figure \ref{fig:FloquetAYDecayDefect}.

After the initial decay, $A_\infty^{Y_{i_0+1}}(\lambda,n)$ stabilizes around a plateau, whose height is given by Eq. (\ref{eqn:plateauheight}) and is much larger than $o(2^{-L})$ for small system sizes. In \cite{Mitra:2023xdo}, we observed similar phenomena for a different deformed Floquet TFI model, where the deformation only included nearest-neighbor Pauli $XX$ interactions. In the following, we will apply the arguments given in \cite{Mitra:2023xdo} to our current deformed Floquet TFI model.

An important ingredient in the explanations for the existence of a plateau is the nature of eigenstates for the Floquet unitary, which changes drastically as the deformation is turned on. When $\lambda = 0$, the Floquet unitary $U_{D,\text{TFI}}$ is quadratic in Majorana fermions within each $\mathbb{Z}_2$ charge sector labeled by $\Omega=\pm 1$ (see Eq. (\ref{eqn:fermionFloquetTFIdefect})). Eigenstates of the Floquet unitary are simple many-particle Fock states $|\Omega,\vec{n}\rangle$, where $n_i$ label the occupation numbers for the fermionic modes after a Bogoliubov transformation. Once $\lambda\neq 0$ is turned on, the four-fermion interactions induce hopping processes between different Fock states. Consequently, eigenstates of the Floquet unitary become delocalized in Fock space, namely the eigenstates are now linear combinations of simple Fock states. 

\onecolumngrid

\insfigsvg{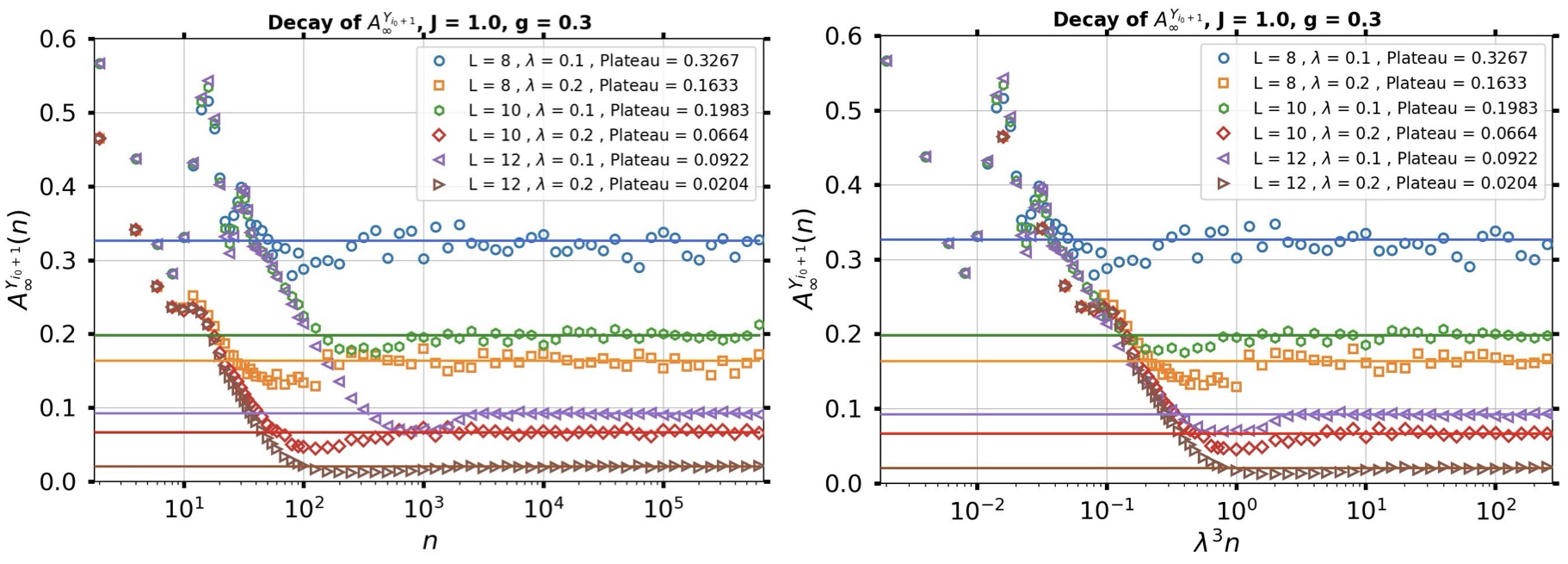}{0.26}{{\it Left Panel} : The infinite-temperature auto-correlation function $A_\infty^{Y_{i_0+1}}$ as defined in Eq. \eqref{eqn:AinftyY}, plotted against discrete time $n$ for $J=1$, $g=0.3$, $T=4$, $L=8,10,12$ and $\lambda=0.1,0.2$. The plateau heights computed using Eq. (\ref{eqn:plateauheight}) are indicated by horizontal solid lines. {\it Right Panel} : The infinite-temperature auto-correlation function $A_\infty^{Y_{i_0+1}}$ plotted against $\lambda^3 n$ for the same set of parameters.}

\twocolumngrid

The degree of this delocalization in Fock space can be quantitatively measured by the inverse participation ratio (IPR), defined as
\begin{equation}\label{eqn:IPR}
\text{IPR}= \frac{1}{2^L}\sum\limits_{\beta_0,\beta}|\langle \beta_0 | \beta\rangle | ^4~,
\end{equation}
where $|\beta_0\rangle$ are eigenstates for the undeformed Floquet unitary and $|\beta\rangle$ are eigenstates for the deformed Floquet unitary. If the eigenstates $|\beta\rangle$ are fully delocalized in Fock space, $|\langle\beta_0|\beta\rangle|\sim 1/\sqrt{2^L}$ and therefore IPR is $o(2^{-L})$. For some numerical examples that we have explored in this manuscript, e.g. $J=1$, $g=0.3$, $T=4$, $\lambda=0.1$ and $L=10$, IPR is computed to be around $0.1502$, which is much larger than $2^{-10}$. This indicates that the eigenstates of the Floquet unitary are only partially delocalized in Fock space for such parameters.

The plateau of the $Y_{i_0+1}$ auto-correlation function is related to Fock-space delocalization in the following physical picture. In a finite-size system, upon increasing $\lambda$, Fock space hopping first induces the eigenstates delocalization in Fock space. Dominantly driven by bulk states scattering processes, the delocalization causes a broadening of quasi-energy levels in the Floquet system. This in turn unlocks scattering processes involving the Majorana zero mode localized at the defect, causing the zero mode to decay. Moreover, the extent of eigenstates delocalization in Fock space governs the plateau height of the auto-correlation function of $Y_{i_0+1}$, which has a non-trivial overlap with the Majorana zero mode.

To corroborate the physical picture, we numerically study the IPR which quantifies the extent of eigenstates delocalization in Fock space, and the normalized plateau height for $A_\infty^{Y_{i_0+1}}(\lambda,n)$ defined as
\begin{equation}\label{eqn:normalizedAYplateau}
\frac{P_{A_\infty^{Y_{i_0+1}}}(\lambda,L)}{P_{A_\infty^{Y_{i_0+1}}}(\lambda=0,L)}~.
\end{equation}
In Figure \ref{fig:J1.0g0.3ARPAYcollect}, we plot these quantities as a function of the deformation strength $\lambda$, for $J=1$, $g=0.3$, $T=4$ and $L=10,12$. Both the IPR and the normalized plateau decrease as $\lambda$ increases, and they both tend to vanish for sufficiently large $\lambda$. Moreover, compared with the normalized plateau, the IPR decreases at a faster rate with respect to $\lambda$. To illustrate the relation between IPR and the normalized plateau, we plot these two quantities directly against each other in Figure \ref{fig:J1.0g0.3IPRPAYcollapse} using the data from Figure \ref{fig:J1.0g0.3ARPAYcollect}. The relation between IPR and the normalized plateau appears to depend on the system size very mildly. For example, a plateau height of around 30 percent of that in the undeformed model, correponds to an IPR value of about 0.1. These numerical results are consistent with the physical picture that the eigenstates delocalization in Fock space governs the plateau height for $A_\infty^{Y_{i_0+1}}(\lambda,n)$.

\insfigsvg{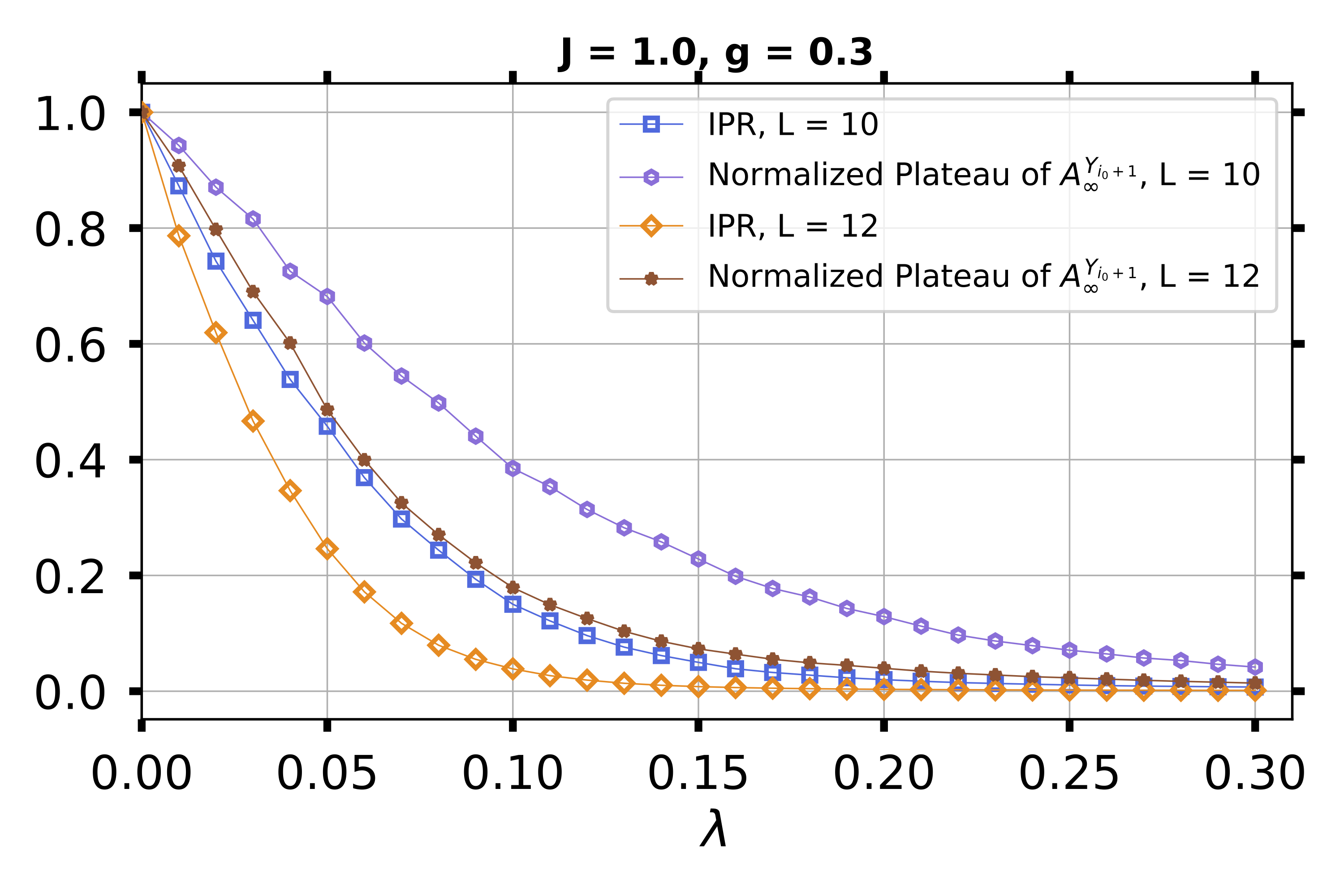}{0.08}{Plots of IPR in Eq. \eqref{eqn:IPR} and normalized plateau of $A_\infty^{Y_{i_0+1}}(\lambda,n)$ in Eq. \eqref{eqn:normalizedAYplateau} as a function of $\lambda$, for $J=1$, $g=0.3$, $T=4$, and $L=10,12$.}

\insfigsvg{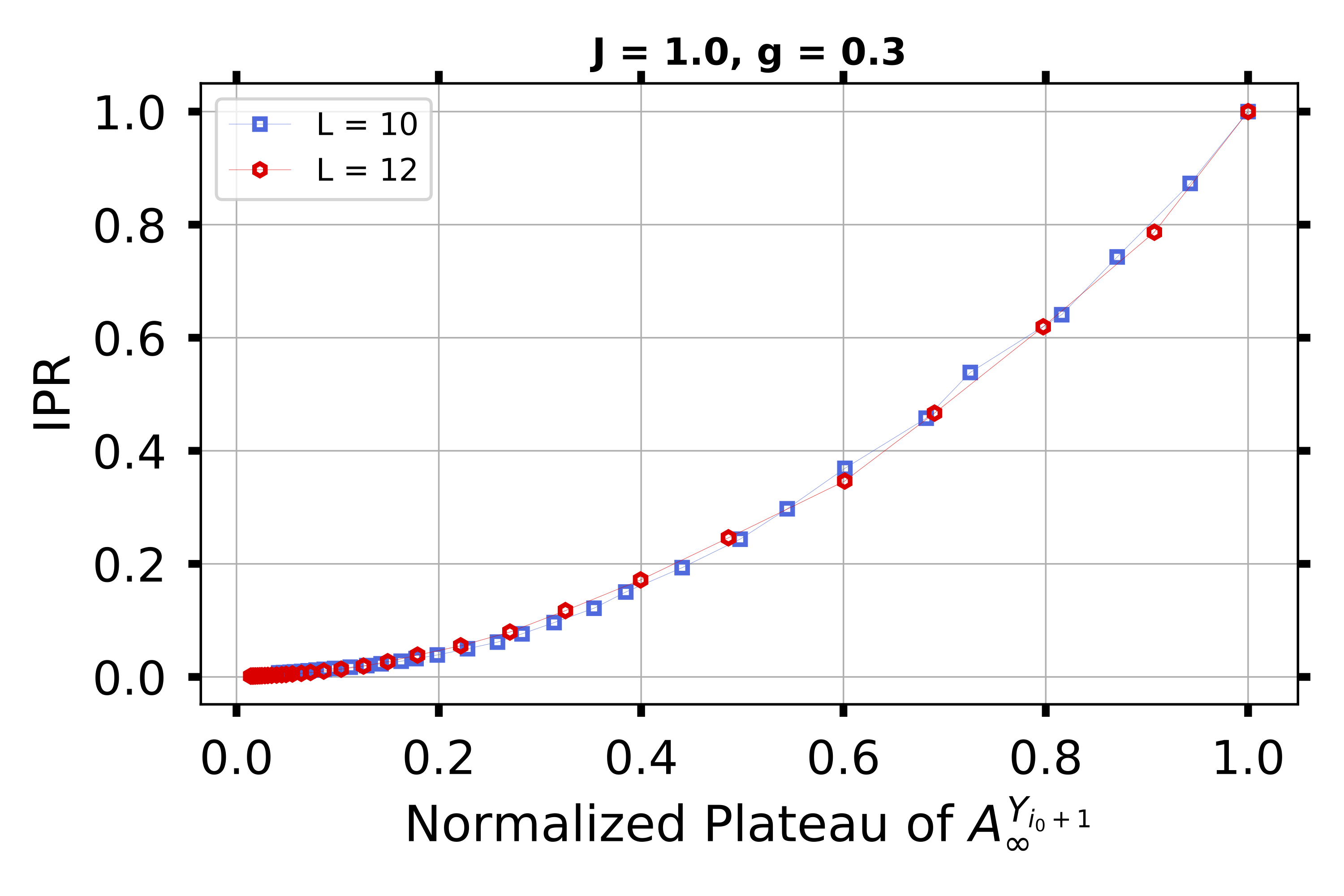}{0.08}{Plot of IPR as defined in Eq. \eqref{eqn:IPR} against the normalized plateau of $A_\infty^{Y_{i_0+1}}(\lambda,n)$ as defined in Eq. \eqref{eqn:normalizedAYplateau}, for $J=1$, $g=0.3$, $T=4$, and $L=10,12$.}

\section{Conclusions}\label{sec:conclusions}

Motivated by constructions of physical quantities with long lifetimes, in this manuscript we aimed to study defect-modified quantum systems under weak integrability breaking deformations. Starting from a defect-modified integrable system, we streamlined the procedure to construct weakly deformed defect-modified extensive local charges, which are expected to possess slower decay behaviors. We also explored the stability of the zero mode localized around the defect, under weak integrability breaking deformations.

We illustrated these ideas using a concrete example: a 1d deformed transverse field Ising model in presence of a Kramers-Wannier duality defect. We investigated the model in the Hamiltonian setting as well as in the periodically-driven Floquet setting. First, we derived the deformed defect Hamiltonian using three distinct methods: half-chain KW transformation, utilization of tools from fusion categories, and defect-modified weak integrability breaking. We then constructed the weakly deformed defect-modified higher charges in the model and studied their slower decay behavior. Additionally, we studied the corresponding deformed Floquet TFI model with a duality defect. In presence of the deformation, the zero mode localized around the defect starts to decay. For small system sizes, the zero mode only decays partially. This effect is quantified by the infinite-temperature auto-correlation function of a local operator overlapping with the zero mode, which stabilizes around a finite plateau at infinite time. 

Even though here we have focused on the particular example of 1d deformed TFI model, the methods demonstrated in this manuscript are expected to apply to more general models. It would be interesting to explore the interplay between defects and weak integrability breaking in additional examples such as the XXZ chain. Additionally, as our setup can be regarded as a lattice version (at leading order) of the $T\overline{T}$ deformation in continuum $(1+1)$-d integrable field theories, it would be interesting to make connections with the study of $T\overline{T}$-deformed theories in presence of boundaries and defects \cite{Jiang:2021jbg,Brizio:2024doe}, which remains largely unexplored.

\bigskip

{\sl Acknowledgements:}\\
We would like to thank Zohar Komargodski for helpful comments on a draft. We would also like to thank David Aasen, Paul Fendley, Zohar Komargodski, Andrew Lucas, Armin Rahmani, Sutapa Samanta, Federica Surace, and Tzu-Chieh Wei for helpful discussions. FY thanks the Simons Center for Geometry and Physics at Stony Brook University for great hospitality during various stages of this work. The work of FY was supported by the U.S. Department of Energy, Office of Basic Energy Sciences, under Contract No. DE-SC0012704. The work of AM was supported by the U.S. National Science Foundation under Grant No. 
NSF DMR-2316598. 

\newpage
\appendix

\section{Detailed calculations of weak integrability breaking deformations}\label{sec:appweakintebreaking}

In this Appendix, we spell out computation details of the weak integrability breaking deformations. We start with the no-defect TFI model discussed in Section \ref{sec:bilocalTFI}. In the following, we will compute the first-order correction to the Hamiltonian $H^{(1)}=-JQ_2^{(1)}$ and see that it reproduces the self-dual deformation in the no-defect TFI model. We will then proceed to compute $Q_1^{(1)}$.  

First, we would need to compute the generalized currents $J_{22}$, $J_{12}$, $J_{21}$ and $J_{11}$. Concretely,
\begin{equation}\label{eqn:comm1}
\begin{split}
\text{i}\left[ Q_2^{(0)},q^{(0)}_{2,j} \right]=&h\left( Y_j Z_{j+1}- Z_{j-1} Y_j\right)\\
&-h\left( Y_{j+1}Z_{j+2}-Z_j Y_{j+1}\right),\\
\text{i}\left[ Q_2^{(0)},q^{(0)}_{1,j} \right]=&\left( -2 Z_{j-1}X_j Z_{j+1}-2 X_j \right)\\
&-\left( -2 Z_{j}X_{j+1} Z_{j+2}-2 X_{j+1} \right) ~.
\end{split}
\end{equation} 
Therefore we have
\begin{equation}
\begin{split}
J^{(0)}_{22,j}&=h\left(Y_jZ_{j+1}- Z_{j-1}Y_j\right)~,\\
J^{(0)}_{12,j}&=-2 Z_{j-1}X_j Z_{j+1}-2 X_j ~.
\end{split}
\end{equation}
Similarly, we can work out the generalized currents $J^{(0)}_{21}$ and $J^{(0)}_{11}$, whose densities are given by
\begin{equation}
\begin{split}
J^{(0)}_{21,j}=&h\left( Y_{j-1}Y_j+Y_jY_{j+1}+Z_{j-1}Z_j+Z_jZ_{j+1}\right)\\
&+2X_j-2Z_{j-1}X_jZ_{j+1}~,\\
J^{(0)}_{11,j}=&2Y_{j-1}X_jZ_{j+1}-2Z_{j-1}X_jY_{j+1}~.
\end{split}
\end{equation}

The first-order deformation to the Hamiltonian, $H^{(1)}$, is given by
\begin{equation}\label{eqn:selfdualdef}
\begin{split}
H^{(1)}=&-\frac{J}{2}\sum\limits_j \{ q^{(0)}_{1,j}, J^{(0)}_{22,j}+J^{(0)}_{22,j+1} \}\\
&+\frac{J}{2}\sum\limits_j \{ q^{(0)}_{2,j}, J^{(0)}_{12,j}+J^{(0)}_{12,j+1} \} ~ ,
\end{split}
\end{equation}
where
\begin{equation}\label{eqn:qJJnodefect}
\begin{split}
\{ q^{(0)}_{1,j}, J^{(0)}_{22,j} \} &=h\left(-2 + 2 Z_{j-1}Z_{j+1} +2 X_j X_{j+1}\right),\\
\{ q^{(0)}_{1,j}, J^{(0)}_{22,j+1} \} &= h\left(2 X_jX_{j+1}  +2 Z_j Z_{j+2}-2\right),\\
\{ q^{(0)}_{2,j}, J^{(0)}_{12,j} \} &=h\left(- 2 Z_{j-1}Z_{j+1}-2 -2 X_j X_{j+1} \right),\\
\{ q^{(0)}_{2,j}, J^{(0)}_{12,j+1} \} &=h\left(-2 X_j X_{j+1} - 2 Z_{j}Z_{j+2}-2\right). 
\end{split}
\end{equation}
Substituing this into Eq. (\ref{eqn:selfdualdef}), we then have 
\begin{equation}\label{eqn:H1nodefect}
H^{(1)} = -4 g \left(\sum\limits_j Z_{j}Z_{j+2} + \sum\limits_j X_j X_{j+1} \right).
\end{equation}
Rescaling the deformation parameter $\lambda$ by a factor of $4g$, we see that $H^{<2}(\lambda)=H^{(0)}+\lambda H^{(1)}$ indeed reproduces the Hamiltonian for the KW self-dual deformed TFI model. We also remark that, to obtain Eq. (\ref{eqn:H1nodefect}), we have relabeled the summation indices for certain terms in Eq. (\ref{eqn:qJJnodefect}), which is justified for the infinite chain setup (as well as for a periodic chain).

We now proceed to compute the deformed charge $Q_1^{<2}(\lambda)=Q_1^{(0)}+\lambda Q_1^{(1)}$, where the undeformed charge $Q_1^{(0)}$ is given in Eq. (\ref{eqn:q1q2nodefect}). From
{\small{\begin{equation*}
\begin{split}
&\{q^{(0)}_{1,j},J^{(0)}_{21,j}\}= -2h Y_{j-1}Z_{j+1}+2h Z_{j-1}Y_{j+1}+4 Z_{j-1}Y_jX_{j+1},\\
&\{q^{(0)}_{1,j},J^{(0)}_{21,j+1}\}= -2h Y_{j}Z_{j+2}+2h Z_{j}Y_{j+2}-4 X_{j}Y_{j+1}Z_{j+2},\\
&\{q^{(0)}_{2,j},J^{(0)}_{11,j}\}= 2h Y_{j-1}Z_{j+1}-2h Z_{j-1}Y_{j+1}-4 Z_{j-1}Y_jX_{j+1},\\
&\{q^{(0)}_{2,j},J^{(0)}_{11,j+1}\}= 2h Y_{j}Z_{j+2}-2h Z_{j}Y_{j+2}+4 X_{j}Y_{j+1}Z_{j+2},
\end{split}
\end{equation*}}} 
together with the rescaling convention of $\lambda \to 4g\lambda$, we obtain
\begin{equation}\label{eqn:Q1nodefect}
\begin{split}
Q_1^{(1)}=&\frac{h}{g}\left(- \sum\limits_j Y_j Z_{j+2}+\sum\limits_j Z_j Y_{j+2} \right) \\
&+ \frac{1}{g}\left( \sum\limits_j Z_jY_{j+1} X_{j+2} - \sum\limits_j X_j Y_{j+1}Z_{j+2} \right),
\end{split}
\end{equation}
with $h=g/J$ in the convention of Eq. (\ref{eqn:HdTFI}).

We now switch to the duality defect twisted TFI model. First of all, as a sanity check, we would like to see that 
\begin{equation}
\left[ Q_2^{D,(0)} , Q_1^{D,(0)} \right]=0 ~.
\end{equation}
Direct computation yields
\begin{equation}
\begin{split}
\left[ Q_2^{D,(0)}, q^{D,(0)}_{1,i_0-1} \right] &= 2\text{i}Z_{i_0-2}X_{i_0-1}Z_{i_0}+2\text{i}X_{i_0-1}\\&-2\text{i}X_{i_0}-2\text{i}Z_{i_0-1}X_{i_0}X_{i_0+1},
\end{split}
\end{equation}

\begin{equation}
\begin{split}
\left[ Q_2^{D,(0)}, q^{D,(0)}_{1,i_0} \right] &=2\text{i}Z_{i_0-1}X_{i_0}X_{i_0+1}-2\text{i}Z_{i_0+1}Z_{i_0+2}\\
&+2\text{i}X_{i_0}+2\text{i}Z_{i_0}Y_{i_0+1}Y_{i_0+2},
\end{split}
\end{equation}

\begin{equation}
\begin{split}
\left[ Q_2^{D,(0)}, q^{D,(0)}_{1,i_0+1} \right] &=-2\text{i}Z_{i_0}Y_{i_0+1}Y_{i_0+2}-2\text{i}hX_{i_0+2}\\
&+2\text{i}Z_{i_0+1}Z_{i_0+2}-2\text{i}hZ_{i_0+1}X_{i_0+2}Z_{i_0+3}.
\end{split}
\end{equation}
Summing these three terms together gives us
\begin{equation}
\begin{split}
&\left(2\text{i}Z_{i_0-2}X_{i_0-1}Z_{i_0} +2\text{i} X_{i_0-1}\right)\\
&-\left( 2\text{i}hZ_{i_0+1}X_{i_0+2}Z_{i_0+3} +2\text{i} hX_{i_0+2} \right)~.
\end{split}
\end{equation}
Comparing with Eq. (\ref{eqn:comm1}), we see that the first two terms are canceled by terms coming from $\left[ Q_2^{D,(0)}, q^{D,(0)}_{1,i_0-2} \right]$. Similarly, the last two terms are canceled by terms coming from $\left[ Q_2^{D,(0)}, q^{D,(0)}_{1,i_0+2} \right]$. In the end, $Q_2^{D,(0)}$ and $Q_1^{D,(0)}$ indeed commute with each other.

Now we are ready to study the deformation generated by the defect-modified bilocal operator $\left[ Q_2^{D,(0)}\big| Q_1^{D,(0)} \right]$. First, we would need to compute the modified generalized current densities, where the modifications happen to sites $j\geq i_0$. For example, the generalized current density $J_{12}^{D,(0)}$ is given by
\begin{equation}
\begin{split}
J_{12,j<i_0}^{D,(0)}&=-2Z_{j-1}X_jZ_{j+1}-2X_j~,\\
J_{12,i_0}^{D,(0)}&=-2 Z_{i_0-1}X_{i_0} X_{i_0+1}-2 X_{i_0}~,\\
J_{12,i_0+1}^{D,(0)}&=2 Z_{i_0} Y_{i_0+1}Y_{i_0+2}-2 Z_{i_0+1}Z_{i_0+2}~,\\
J_{12,j>i_0+1}^{D,(0)}&=-2hZ_{j-1}X_jZ_{j+1}-2hX_j~.
\end{split}
\end{equation}
Similarly the generalized current density $J_{22}^{D,(0)}$ is
\begin{equation}
\begin{split}
J_{22,i_0}^{D,(0)}&=-hZ_{i_0-1}Y_{i_0}+hY_{i_0}X_{i_0+1}~,\\
J_{22,i_0+1}^{D,(0)}&=2hZ_{i_0}Y_{i_0+1}Z_{i_0+2}~,\\
J_{22,j\neq i_0,i_0+1}^{D,(0)}&=J_{22,j}^{(0)}~.
\end{split}
\end{equation}

Next we compute the following anti-commutators
{\small{\begin{equation*}
\begin{split}
\{ q_{1,i_0-1}^{D,(0)},J_{22,i_0}^{D,(0)} \}& = 2h X_{i_0-1}X_{i_0}+2h Z_{i_0-1} X_{i_0+1}-2h~,\\
\{ q_{1,i_0}^{D,(0)},J_{22,i_0}^{D,(0)} \}& = -2h+2h Z_{i_0-1}X_{i_0+1}+2h X_{i_0}Z_{i_0+1} Z_{i_0+2},\\
\{ q_{1,i_0}^{D,(0)},J_{22,i_0+1}^{D,(0)} \}& = 4hX_{i_0}Z_{i_0+1}Z_{i_0+2}-4h~,\\
\{ q_{1,i_0+1}^{D,(0)},J_{22,i_0+1}^{D,(0)} \}& =   4h Z_{i_0}X_{i_0+1}X_{i_0+2}~,\\
\{ q_{1,i_0+1}^{D,(0)},J_{22,i_0+2}^{D,(0)} \}& = 2hZ_{i_0+1}Z_{i_0+3}-2h~,
\end{split}
\end{equation*}}}
together with
{\small{\begin{equation*}
\begin{split}
\{ q_{2,i_0-1}^{D,(0)},J_{12,i_0}^{D,(0)} \}&=-2h X_{i_0-1}X_{i_0}-2h Z_{i_0-1}X_{i_0+1}-2h~,\\
\{ q_{2,i_0}^{D,(0)},J_{12,i_0}^{D,(0)} \}&=-2h Z_{i_0-1}X_{i_0+1}-2h~,\\
\{ q_{2,i_0}^{D,(0)},J_{12,i_0+1}^{D,(0)} \}&=-2h X_{i_0}Z_{i_0+1}Z_{i_0+2}~,\\
\{ q_{2,i_0+1}^{D,(0)},J_{12,i_0+1}^{D,(0)} \}&=-4h Z_{i_0}X_{i_0+1}X_{i_0+2}-4h~,\\
\{ q_{2,i_0+1}^{D,(0)},J_{12,i_0+2}^{D,(0)} \}&=-2h Z_{i_0+1}Z_{i_0+3}-2h~.
\end{split}
\end{equation*}}}

Putting everything together, the deformation to the defect Hamiltonian in integrable TFI model is 
\begin{equation}
\begin{split}
H^{D,(1)}&=-\frac{J}{2}\sum\limits_j \{q_{1,j}^{D,(0)}, J_{22,j}^{D,(0)}+J_{22,j+1}^{D,(0)} \}\\
&+\frac{J}{2}\sum\limits_j \{q_{2,j}^{D,(0)}, J_{12,j}^{D,(0)}+J_{12,j+1}^{D,(0)} \}\\
=&-4g\bigg(\sum\limits_{\substack{j\neq i_0-1,\\j\neq i_0}}Z_j Z_{j+2}+\sum\limits_{\substack{j\neq i_0,\\j\neq i_0+1}}X_j X_{j+1}\\
&+Z_{i_0-1}X_{i_0+1}+X_{i_0}Z_{i_0+1}Z_{i_0+2}+Z_{i_0}X_{i_0+1}X_{i_0+2}\bigg).
\end{split}
\end{equation}
Rescaling $\lambda \to 4g\lambda$ as before, this reproduces the deformation terms, including the extra local modifications, in the defect Hamiltonian Eq. (\ref{eqn:H_DdTFI1}) for self-dual duality-twisted TFI model.

Next, we will compute the deformed charge $Q_1^{D,<2}(\lambda)$. From Eq. (\ref{eqn:q1q2nodefect}) and Eq. (\ref{eqn:q1defect}), the undeformed duality-twisted charge is given by
\begin{equation}
\begin{split}
Q_1^{D,(0)}&=\sum\limits_{\substack{j,\\ j\neq i_0,i_0+1}} \left(-Y_jZ_{j+1}+Z_j Y_{j+1}\right)\\
&-Y_{i_0}X_{i_0+1}-Z_{i_0}Y_{i_0+1}Z_{i_0+2}+Z_{i_0+1}Y_{i_0+2}~.
\end{split}
\end{equation}
To compute $Q_1^{D,(1)}$, we first evaluate the defect-modified generalized current densities. Concretely,
{\small{\begin{equation*}
\begin{split}
J_{21,j<i_0}^{D,(0)}&=J_{21,j}^{(0)}~,\\
J_{21,i_0}^{D,(0)}&=h(Y_{i_0-1}Y_{i_0}+Z_{i_0-1}Z_{i_0}+Z_{i_0}X_{i_0+1}\\
&-Y_{i_0}Y_{i_0+1}Z_{i_0+2})+2X_{i_0}-2Z_{i_0-1}X_{i_0}X_{i_0+1}~,\\
J_{21,i_0+1}^{D,(0)}&=2hZ_{i_0}X_{i_0+1}-2hY_{i_0}Y_{i_0+1}Z_{i_0+2}\\
&+2Z_{i_0}Y_{i_0+1}Y_{i_0+2}+2Z_{i_0+1}Z_{i_0+2}~,\\
J_{21,i_0+2}^{D,(0)}&=Z_{i_0}Y_{i_0+1}Y_{i_0+2}+Z_{i_0+1}Z_{i_0+2}+Z_{i_0+2}Z_{i_0+3}\\
&+Y_{i_0+2}Y_{i_0+3}+2hX_{i_0+2}-2hZ_{i_0+1}X_{i_0+2}Z_{i_0+3}~,\\
J_{21,j>i_0+2}^{D,(0)}&=Z_{j-1}Z_j+Z_jZ_{j+1}+Y_{j-1}Y_j+Y_jY_{j+1}\\
&+2hX_j -2h Z_{j-1}X_jZ_{j+1}~.
\end{split}
\end{equation*}}}
Additionally we have
{\small{\begin{equation*}
\begin{split}
J_{11,j<i_0}^{D,(0)}&=J_{11,j}^{(0)}~, ~J_{11,j>i_0+2}^{D,(0)}=J_{11,j}^{(0)}~,\\
J_{11,i_0}^{D,(0)}&=2Y_{i_0-1}X_{i_0}X_{i_0+1}+2Z_{i_0-1}X_{i_0}Y_{i_0+1}Z_{i_0+2}~,\\
J_{11,i_0+1}^{D,(0)}&=-2Y_{i_0}Y_{i_0+1}Y_{i_0+2}+2Z_{i_0}Y_{i_0+1}X_{i_0+2}Z_{i_0+3}~,\\
J_{11,i_0+2}^{D,(0)}&=2Z_{i_0}Y_{i_0+1}X_{i_0+2}Z_{i_0+3}~-2Z_{i_0+1}X_{i_0+2}Y_{i_0+3}.
\end{split}
\end{equation*}}}

Using these modified generalized currents, the $o(\lambda)$ deformation in presence of a duality interface, $Q_1^{D,(1)}$, is given by
{\small{\begin{equation}\label{eqn:Q11defect}
\begin{split}
&Q_1^{D,(1)}=\frac{1}{J}\sum\limits_{j\leq i_0-2} (Z_jY_{j+2}-Y_jZ_{j+2})\\
&+\frac{1}{g}\sum\limits_{j\leq i_0-2} (Z_jY_{j+1}X_{j+2}-X_jY_{j+1}Z_{j+2})\\
&+\frac{1}{g}\sum\limits_{j\geq i_0+2} (Z_jY_{j+2}-Y_jZ_{j+2})\\
&+\frac{1}{J}\sum\limits_{j\geq i_0+2} (Z_jY_{j+1}X_{j+2}-X_jY_{j+1}Z_{j+2})\\
&+\frac{1}{g}Z_{i_0+1}Y_{i_0+3}+\frac{1}{J}Z_{i_0+1}Y_{i_0+2}X_{i_0+3}-\frac{1}{J}Y_{i_0-1}X_{i_0+1}\\
&-\frac{1}{g}X_{i_0-1}Y_{i_0}X_{i_0+1}+\frac{1}{g}X_{i_0}Z_{i_0+1}Y_{i_0+2}-\frac{1}{g}Z_{i_0}Y_{i_0+1}Z_{i_0+3}\\
&-\frac{1}{J} Z_{i_0-1}Y_{i_0+1}Z_{i_0+2}-\frac{1}{J}Y_{i_0}X_{i_0+1}X_{i_0+2}\\
&+\frac{1}{g}Z_{i_0-1}Y_{i_0}Z_{i_0+1}Z_{i_0+2}-\frac{1}{J}Z_{i_0}X_{i_0+1}Y_{i_0+2}Z_{i_0+3}
\end{split}
\end{equation}}}
where we have again applied the rescaling convention of $\lambda\to 4g\lambda$. Notice that, $Q_1^{D,(1)}$ in Eq. (\ref{eqn:Q11defect}) can also be constructed from performing half-space KW transformation on $Q_1^{(1)}$ (Eq. (\ref{eqn:Q1nodefect})) in the no-defect case. We have checked that the results do agree from these two methods.

\section{Decay behavior analysis for the deformed charges at early times}\label{sec:decayrate}

In this Appendix, we provide some details of the perturbative analysis for the decay of deformed charges at early times. Concretely, we consider the time derivative of $A^Q_\infty(\lambda,t)$:
{\small{\begin{equation}
\begin{split}
\frac{d}{dt}A_\infty^Q(\lambda,t)&=\frac{1}{\mathcal{N}}\frac{d}{dt}\text{Tr}\left( \text{e}^{\text{i}H(\lambda)t}Q(\lambda)\text{e}^{-\text{i}H(\lambda)t}Q(\lambda) \right)\\
&=\frac{\text{i}}{\mathcal{N}}\text{Tr}\left( \text{e}^{\text{i}H(\lambda)t}\left[H(\lambda),Q(\lambda)\right]\text{e}^{-\text{i}H(\lambda)t}Q(\lambda) \right)
\end{split}
\end{equation}}}
Recall that $\left[H(\lambda),Q(\lambda)\right]=\lambda^2[H^{(1)},Q^{(1)}]$, so the formal leading order dependence is at $o(\lambda^2)$. We will show below that this formal leading order vanishes for an infinite chain.

The time evolution of any generic operator $\mathcal{O}$ in the Heisenberg picture is given by
\begin{equation}
\begin{split}
\text{e}^{\text{i}H(\lambda)t}\mathcal{O}\text{e}^{-\text{i}H(\lambda)t}&=\overline{T}\left\{ \text{exp}\left( \text{i}\lambda\int_0^t dt'H^{(1)}_0(t') \right) \right\}\mathcal{O}_0(t)\\
&\times T\left\{ \text{exp}\left( -\text{i}\lambda\int_0^t dt'H^{(1)}_0(t') \right) \right\}
\end{split},
\end{equation}
where
\begin{equation}
\begin{split}
\mathcal{O}_0(t)&:=\text{e}^{\text{i}H^{(0)}t}\mathcal{O} \text{e}^{-\text{i}H^{(0)}t}~,\\
H_0^{(1)}(t)&:= \text{e}^{\text{i}H^{(0)}t}H^{(1)} \text{e}^{-\text{i}H^{(0)}t}~,
\end{split}
\end{equation}
and $T$ and $\overline{T}$ are the time ordering and anti-time ordering operators respectively.

We can then compute the time derivative perturbatively in $\lambda$, here up to $o(\lambda^3)$. We have
\begin{equation}
\frac{d}{dt}A^Q_\infty(\lambda,t)=\frac{1}{\mathcal{N}}(c_2\lambda^2+c_3\lambda^3+o(\lambda^4)),
\end{equation}
with
\begin{equation}\label{eqn:c2eqn}
\begin{split}
c_2&=\text{i}\text{Tr}\left( \text{e}^{\text{i}H_0t}\left[ H^{(1)},Q^{(1)}\right] \text{e}^{-\text{i}H_0t}Q^{(0)}\right)\\
&=\text{i}\text{Tr}\left( \left[ H^{(1)},Q^{(1)}\right] Q^{(0)} \right)~,
\end{split}
\end{equation}
and $c_3$ is given by
{\small{\begin{equation}
\begin{split}
&-\text{Tr}\left( \int_0^t dt'\left[ H^{(1)}_0(t'), \left(\left[H^{(1)},Q^{(1)} \right] \right)_0(t)\right] Q^{(0)}\right)\\
&+\text{i}\text{Tr}\left( \text{e}^{\text{i}H^{(0)}t}\left[ H^{(1)},Q^{(1)}\right]\text{e}^{-\text{i}H^{(0)}t} Q^{(1)}\right)\\
=&-\int_0^t dt' \text{Tr}\left( \text{e}^{\text{i}H_0(t-t')} \left[H^{(1)},Q^{(1)} \right]\text{e}^{-\text{i}H_0(t-t')} \left[Q^{(0)},H^{(1)} \right]\right)\\
&+\text{i}\text{Tr}\left( \text{e}^{\text{i}H^{(0)}t}\left[ H^{(1)},Q^{(1)}\right]\text{e}^{-\text{i}H^{(0)}t} Q^{(1)}\right)\\
=&-\int_0^t dt' \text{Tr}\left( \text{e}^{\text{i}H_0t'} \left[H^{(1)},Q^{(1)} \right]\text{e}^{-\text{i}H_0t'} \left[Q^{(0)},H^{(1)} \right]\right)\\
&+\text{i}\text{Tr}\left( \text{e}^{\text{i}H^{(0)}t}\left[ H^{(1)},Q^{(1)}\right]\text{e}^{-\text{i}H^{(0)}t} Q^{(1)}\right).
\end{split}
\end{equation}}}

To show the formal vanishing of $c_2$, notice that
{\small{\begin{equation}\label{eqn:h1q1}
\begin{split}
\left[ H^{(1)},Q^{(1)} \right]&= X^{(0)}Q^{(0)}X^{(0)}H^{(0)}-Q^{(0)}X^{(0)}X^{(0)}H^{(0)}\\
&+Q^{(0)}X^{(0)}H^{(0)}X^{(0)}
-X^{(0)}H^{(0)}X^{(0)}Q^{(0)}\\
&+H^{(0)}X^{(0)}X^{(0)}Q^{(0)}-H^{(0)}X^{(0)}Q^{(0)}X^{(0)} ~.
\end{split}
\end{equation}}}
Substituting this into Eq.~(\ref{eqn:c2eqn}) and using the cyclic property of the trace yields $c_2=0$. This means that the early decay behavior of $A_\infty^Q(\lambda,t)$ has at least an $o(\lambda^3)$ dependence on the deformation strength.

The expression for $\left[H^{(1)},Q^{(1)}\right]$ in Eq.~(\ref{eqn:h1q1}) is valid given that the $0$-th order generator $X^{(0)}$ is well-defined. This is certainly true for the setup of an infinite chain, but raises potential issues for a finite periodic chain. While we do not have an analytical proof for the case of a finite periodic chain, our numerical analysis in Section \ref{sec:bilocalTFI} and Section \ref{sec:weakintegrabilitybreakingKW} does support an $o(\lambda^3)$ dependence of $A_\infty^Q(\lambda,t)$ at early times.  

\section{The ordering convention for Floquet unitaries}\label{sec:FloOrdering}

In this Appendix, we spell out the details for the ordering of non-commuting terms in the Floquet unitary. We will start with the undeformed Floquet TFI model, where the Floquet unitary is given in Eq. (\ref{eqn:FloquetTFI}). The duality defect twisted Floquet TFI model was studied in \cite{Tan22}, where the Floquet unitary in Eq. (\ref{eqn:FloquetTFIdefect}) was derived using the language of face models in \cite{Fendley16}. Here we will reinterpret the ordering prescription for the Floquet unitary using a half-space sequential KW transformation. 

We consider a chain with a total of $L$ sites, where we take $L$ large and ignore the details at the boundaries. The half-space KW transformation, denoted as $\widehat{D}_{i_0}$, acts on the local Hamiltonian terms in the following way
\begin{equation}
\begin{split}
\widehat{D}_{i_0}X_j &= Z_j Z_{j+1}\widehat{D}_{i_0}~, ~ j\geq i_0+1~,\\
\widehat{D}_{i_0}Z_jZ_{j+1}&=X_{j+1}\widehat{D}_{i_0}~,~ j\geq i_0+1~,\\
\widehat{D}_{i_0}Z_{i_0}Z_{i_0+1}&=Z_{i_0}X_{i_0+1}\widehat{D}_{i_0}~.
\end{split}
\end{equation}
Applying $\widehat{D}_{i_0}$ on the Floquet unitary $U_{\text{TFI}}$ in Eq. (\ref{eqn:FloquetTFI}), we obtain
\begin{equation}
\widehat{D}_{i_0}U_{\text{TFI}}=U'_{\text{TFI}}\widehat{D}_{i_0}~,
\end{equation}
with
\begin{equation}\label{eqn:UprimeTFI}
\begin{split}
U'_{\text{TFI}}&=\left(\prod\limits_{j=1}^{i_0}\text{e}^{-\text{i}u_j^b X_j}\right)\left(\prod\limits_{j=i_0+1}^L\text{e}^{-\text{i}u_j^b Z_j Z_{j+1}}\right)\\
&\times \left( \prod\limits_{j=1}^{i_0-1}\text{e}^{-\text{i}u_j^a Z_j Z_{j+1}} \right)\text{e}^{-\text{i}u^a_{i_0}Z_{i_0}X_{i_0+1}}\\
&\times \left(\prod\limits_{j=i_0+1}^L  \text{e}^{-\text{i}u_j^a X_{j+1}}\right)~.
\end{split}
\end{equation}
Clearly, $U'_{\text{TFI}}$ is not the same as $U_{D,\text{TFI}}$ in Eq. (\ref{eqn:FloquetTFIdefect}), and it can't be applied on a periodic chain. In \cite{Tan22}, an ordering prescription was given in the language of face models, which translates into exchanging $Z_jZ_{j+1}$ with $X_{j+1}$ for $j\geq i_0+1$ in $U'_{\text{TFI}}$. Concretely, doing so we obtain the following Floquet unitary
{\small{\begin{equation}\label{eqn:UprimeDTFI}
\begin{split}
U'_{D,\text{TFI}}&=\left(\prod_{\substack{j=1,\\ j\neq i_0+1}}^{L+1}\text{e}^{-\text{i}u_j^b X_j}\right)\left(\prod_{j=1}^{i_0-1}\text{e}^{-\text{i}u_j^a Z_jZ_{j+1}}\right)\\
&\times\text{e}^{-\text{i}u_{i_0}^a Z_{i_0}X_{i_0+1}}\left(\prod_{j=i_0+1}^{L}\text{e}^{-\text{i}u_j^a Z_jZ_{j+1}}\right)\\
=&\left(\prod\limits_{\substack{j=1,\\ j\neq i_0+1}}^{L+1} \text{e}^{-\text{i}u_j^b X_j}\right)\times \text{e}^{-\text{i}u^a_{i_0}Z_{i_0}X_{i_0+1}}\\
&\times\left(\prod\limits_{\substack{j=1,\\j\neq i_0}}^L \text{e}^{-\text{i}u_j^a  Z_j Z_{j+1}}\right),
\end{split}
\end{equation}}}
where we have relabeled $u_j^b$ for $j> i_0+1$. In particular, this recovers $U_{D,
\text{TFI}}$ in Eq. (\ref{eqn:FloquetTFIdefect}) up to a boundary term.

We can now apply this prescription to the deformed Floquet TFI model. Starting from the no-defect Floquet unitary in Eq. (\ref{eqn:dFloquetTFI}) and apply $\widehat{D}_{i_0}$, we have
\begin{equation}
\widehat{D}_{i_0} U_{\text{dTFI}} = U'_{\text{dTFI}} \widehat{D}_{i_0}~,
\end{equation}
where 
{\small{\begin{equation}
\begin{split}
U'_{\text{dTFI}}&=\left(\prod\limits_{j=1}^{i_0-1}\text{e}^{-\text{i}v_j^b X_j X_{j+1}}\right)\text{e}^{-\text{i}v_{i_0}^bX_{i_0}Z_{i_0+1}Z_{i_0+2}}\\
&\times\left(\prod\limits_{j=i_0+1}^{L}\text{e}^{-\text{i}v_j^b Z_j Z_{j+2}}\right)\left(\prod\limits_{j=1}^{i_0-2}\text{e}^{-\text{i}v_j^a Z_j Z_{j+2}}\right)\\
&\times \text{e}^{-\text{i}v_{i_0-1}^aZ_{i_0-1}X_{i_0+1}}\text{e}^{-\text{i}v_{i_0}^aZ_{i_0}X_{i_0+1}X_{i_0+2}}\\
&\left(\prod\limits_{j=i_0+1}^{L}\text{e}^{-\text{i}v_j^a X_{j+1} X_{j+2}}\right) U'_{\text{TFI}}~,
\end{split}
\end{equation}}}
with $U'_{\text{TFI}}$ as given in Eq. (\ref{eqn:UprimeTFI}). 

Similar to the case of the undeformed Floquet TFI model, to obtain the duality defect twisted Floquet unitary for the deformed TFI model, we exchange $Z_jZ_{j+1}$ with $X_{j+1}$ (therefore we also exchange $Z_jZ_{j+2}$ with $X_{j+1}X_{j+2}$) for $j\geq i_0+1$. In the end, we have
{\small{\begin{equation}
\begin{split}
U'_{D,\text{dTFI}}&=\left(\prod\limits_{j=1}^{i_0-1}\text{e}^{-\text{i}v_j^b X_j X_{j+1}}\right)\text{e}^{-\text{i}v_{i_0}^bX_{i_0}Z_{i_0+1}Z_{i_0+2}}\\
&\times \left(\prod\limits_{j=i_0+1}^{L}\text{e}^{-\text{i}v_j^b X_{j+1} X_{j+2}}\right)\left(\prod\limits_{j=1}^{i_0-2}\text{e}^{-\text{i}v_j^a Z_j Z_{j+2}}\right)\\
&\times \text{e}^{-\text{i}v_{i_0-1}^aZ_{i_0-1}X_{i_0+1}}\text{e}^{-\text{i}v_{i_0}^aZ_{i_0}X_{i_0+1}X_{i_0+2}}\\
&\times \left(\prod\limits_{j=i_0+1}^{L}\text{e}^{-\text{i}v_j^a Z_j Z_{j+2}}\right)U'_{D,\text{TFI}}~.
\end{split}
\end{equation}}}
After reshuffling commuting terms and relabeling of $v_j^b$ for $j> i_0+1$, we again recover $U_{D,\text{dTFI}}$ in Eq. (\ref{eqn:dFloquetDTFI}) up to a boundary term.

\newpage

%\nocite{apsrev42Control}
%\bibliography{biblio}
%\bibliographystyle{apsrev4-2}


\end{document}